\documentclass[11pt]{article}
\usepackage{graphicx, amsmath, amssymb}
\usepackage{verse}
\usepackage{textgreek}
\usepackage[textsize=tiny]{todonotes}
\usepackage{comment}
\usepackage[shortlabels]{enumitem}

\usepackage[margin=1.3in]{geometry}

\usepackage[greek, english]{babel}

\usepackage[style=ext-authoryear-comp,maxcitenames=2,uniquename=false,backend=biber,uniquelist=false,articlein=false]{biblatex}
\DeclareNameAlias{sortname}{family-given}

\usepackage{hyperref, ebgaramond}
\usepackage[usenames,dvipsnames]{xcolor}

\setlist[itemize]{
    parsep=0pt,
    listparindent=\parindent
}

\hypersetup{
	colorlinks=true,            
        breaklinks=true,
	linkcolor=Black,
	citecolor=MidnightBlue,
	urlcolor=MidnightBlue            
 }

 \makeatletter
\DeclareCiteCommand{\citeyear}
  {\usebibmacro{prenote}}
  {\printtext[bibhyperref]{\usebibmacro{citeyear}}}
  {\multicitedelim}
  {\usebibmacro{postnote}}
\makeatother

\title{Dark Matters of Principle: Principles in the Dark Matter Paradigm}
\author{Patrick M.\ Duerr\footnote{HPS Department, University of Cambridge, UK; patrick-duerr@gmx.de}~ \& James Read\footnote{Faculty of Philosophy, University of Oxford, UK \& Pembroke College; james.read@philosophy.ox.ac.uk}}
\date{}

\begin{document}

\maketitle

\begin{abstract}
\fontsize{10}{12}\selectfont

This paper re-examines the Dark Matter (DM) problem through the lens of physical principles. We proceed in two steps. Taking inspiration from some strands of neo-Kantianism, we first develop a functionalist account and taxonomy of principles. We propose to define principles in terms of their intended functions or purposes (broadly grouped into ``constitutive’’ and ``regulative’’ ones). Their argumentative force and the warrant scientists have for adopting such principles boil down to context-sensitive and often tentative reasons for believing them adequate for those purposes. In research contexts in which novel theories or models are sought which push the boundaries of knowledge into new domains not (yet) under good epistemic control, reliance on well-motivated principles as principal building blocks is a historically widespread, natural, and methodologically prudent research strategy for cautious innovation---what we’ll dub ``principled inquiry'’. In a second step, we apply this methodological machinery to the DM problem. Through detailed analyses of four mainstream DM approaches---supersymmetric WIMPs, axions, sterile neutrinos, and primordial black holes---we substantiate four claims. \textit{(C1)} The DM paradigm, which comprises those four dominant DM candidates, is formed by the framework of possibilities, circumscribed by a \textit{shared cluster} of regulative and constitutive principles of gravitational and non-gravitational physics. \textit{(C2)} Besides those common principles, the four DM models also depart from \textit{other} principles of orthodox physics in distinct ways. \textit{(C3)} They do so to different degrees---displaying different forms and extents of ``minimal mutilation” with respect to some of the principles of established non-gravitational physics (viz., of the Standard Model of Particle Physics). As a research strategy with normative thrust, principled inquiry furnishes natural criteria for comparatively assessing proposals within the paradigm. In particular, it suggests that today research on sterile neutrinos or axions be prioritised over research on WIMPs or primordial black holes. \textit{(C4)} The guidelines of principled inquiry clarify the sense in which the DM problem constitutes an ever-more disconcerting crisis---rather than merely a garden-variety of evidential underdetermination: worries are growing that the principles on which DM researchers have routinely and rationally relied in constructing viable DM models may have run out of heuristic steam. Pressure is mounting to ``leave no stone unturned’’ \parencite{BertoneTait2018}: scientists ought to increasingly explore ideas that stray more radically from the cautious explorations of principled inquiry.

\medskip


\pagebreak





\end{abstract}

\tableofcontents

\section{Introduction}\label{sec:intro}


Multiple crises afflict contemporary physics. To name only the most prominent: the Dark Energy problem (e.g.,\ \cite[pp.\ 5, 10]{Bousso_2007}), the lack of empirical evidence for supersymmetry in high-energy physics (e.g.,\ \cite{LykkenSpiropulu2014}), the Dark Matter problem (e.g., \cite{Kroupa2012}, \cite[p.\ 51]{BertoneTait2018}), and the deadlock and stagnation in quantum gravity research (e.g.,\ \cite{Smolin2006Trouble}). They have all been proclaimed major crises in today's fundamental physics.   

It would be presumptuous for a philosophical paper to even purport to contribute substantially to a \textit{solution} to these crises. A more modest---and hopefully more realistic---goal is to shed light on the nature of those profound quandaries \textit{qua crises}. What exactly do such crises amount to? That is, what makes certain knotty puzzles qualify as crises? How do they emerge? What do crises portend? Do they invariably live up to their etymology---as turning points (and if so, turning points in which regard)?\footnote{Talk of `crisis' is likely to elicit associations with Kuhn's account of extraordinary and revolutionary science (see, e.g.,\ \textcite[chs.\ 6, 7]{HoyningenHuene1993Reconstructing}). We forgo a comparison here since it concerns the most controversial aspects of Kuhn's model of theory dynamics.}

Our focus will be on the crisis associated with the Dark Matter (DM) problem.\footnote{The reader might wonder: why, then, pick the \textit{DM} crisis---instead of the Dark Energy crisis or the quantum gravity crisis? Our choice isn’t supposed to be related to the significance of those other crises. Rather, the DM problem exemplarily resembles the deadlock, eloquently described by Poincaré: a robust phenomenon proves to be a durable problem, obstinately resisting a satisfactory resolution within the confines of principle-guided problem-solving (the research strategy that, in \S\ref{sec:principles}, we’ll dub `principled inquiry'). This predicament precipitates a state of perplexity amongst researchers: that a powerful and common method reached its heuristic limits; a turning point---a `crisis' now also in the literal sense---in their research strategies seems called for.

The crisis in \textit{quantum gravity} is different in two key regards. (i) It’s not so much concerned with unsuccessfully tackling a robust empirical phenomenon, rather than a set of abstract, conceptual-theoretical goals (see, e.g.,\ \textcite{Crowther2025}); moreover, opinions tend to be divided over their cogency and interpretation. (ii) The state of affairs in quantum gravity that tends to be deplored (e.g.,\ \textcite{Hossenfelder2018}) is one of theoretical stagnation and of lack of empirical predictions/tests and salient data that might further guide developments. 

The crisis surrounding the \textit{Dark Energy} problem is likewise different. (i) Whereas in the case of DM a host of independent, diverse constraints render the anomaly in question robust (\S\ref{sec:DMcrisis}), there is a sense in which this \textit{doesn’t} apply to the Dark Energy case \parencite{Durrer2011DarkEnergy}. (ii) What exactly the Dark Energy problem consists in turns out to be controversial---and in particular the extent to which it counts as a problem or anomaly \textit{sensu stricto} (see \textcite{Koberinski2021VacuumEnergy, Schneider2020CosmoConst, KoberinskiSmeenk2023CosmoConst}).}
It denotes the mismatch between the gravitational effects in astrophysical and cosmological observations and the mass that can be accounted for by luminous matter alone. Galactic rotation curves, gravitational lensing, cosmic microwave background measurements, and large-scale structure formation all suggest that approximately 85\% of the universe’s matter is `dark'---that is, interacts primarily via gravity without emitting or absorbing light.

This mismatch has stubbornly resisted a satisfactory resolution. It would be wrong, however, to attribute the crisis to a dearth of seemingly viable ideas. What makes the Dark Matter problem so vexing is rather the opposite: while some proposals have been ruled out, a formidable plenitude lingers on, with no shortage of new ideas. At present, discriminating amongst them on empirical grounds doesn’t seem in the offing. Is then the crisis associated with the Dark Matter problem \textit{merely} an instance of (as scientists would hope: transient) underdetermination of theory by data?  

Indeed, extant philosophical analyses (e.g., \textcite{Vanderburgh2014_InterpretiveRole, MassimiPeacock2014_DarkMatterDE, SmeenkEllis2017_PhilosophyOfCosmology, MartensKing2023_DoingMoreWithLess, DuerrWolf2023_MethodologicalMOND, Antoniou2025_WhyDarkMatterSuperseded, FerreiraWolfRead2025_SpectreUnderdetermination})
have largely examined the DM problem along those lines.\footnote{In particular, theory virtues have figured centrally in them. Following, e.g.,\ \textcite{Kuhn1977EssentialTension} (cf.\ \textcite{DuerrFischer2025}  for a systematic elaboration), this is unsurprising: theory virtues are routinely invoked in the hopes of breaking underdetermination. \textcite{Tulodziecki2025_DataUnderdetermined, Tulodziecki2025_UnderdeterminationVirtues}  has cast doubt on that hope.} We don’t impugn the significance of underdetermination: it \textit{is} integral to the DM problem. Yet, something peculiar to the underdetermination renders the DM problem so irksome---or so we contend.\footnote{Ironically, \textit{insofar} as transient underdetermination is a pervasive phenomenon, as some argue (e.g.,\ \textcite{sep-scientific-underdetermination}), the increasing sense of despair in the DM community would appear puzzling: whence such pronounced frustration if underdetermination is supposed to be so common?} 

Let’s take a cue from Einstein. Referring to underdetermination (and fully endorsing theory-ladenness of observations), he writes (\citeyear{Einstein1918MotiveDesForschens}
p.\ 31 translation by Don Howard): ``[i]n this state of methodological uncertainty one can think that arbitrarily many, in themselves equally justified systems of theoretical principles were possible; and this opinion is, in principle, certainly correct. But the development of physics has shown that of all the conceivable theoretical constructions a single one has, at any given time, proved itself unconditionally superior to all others. No one who has really gone deeply into the subject will deny that, in practice, the world of perceptions determines the theoretical system unambiguously, even though no logical path leads from the perceptions to the basic principles of the theory.” The allusion to principles isn't casual. As \textcite[p.\ 2]{Giovanelli} has pointed out, Einstein jokingly self-labelled himself as ```a principle-pincher' (\emph{Prinzipienfuchser}), ready to squeeze as much as possible from few fundamental principles, rather than a profligate virtuoso, squandering his calculation mastery in trifling puzzle solving”.

Another philosopher-scientist, revered by Einstein, extolled the role of principles in a similar vein. Likewise accepting underdetermination, Poincaré encapsulated his reflections, intriguingly, in a paper the year before Einstein’s \textit{annus mirabilis}. Elaborating a connection between crises and principles, \textcite{poincare1904etat} descries ``signs of a profound crisis, suggesting that a major transformation is imminent” (p.\ 302, our translation) and offers ``un petit diagnostic” (p.\ 302).

Scientific inquiry, according to Poincaré, is and ought to be guided by principles. Poincaré enumerates ``five or six” principal ones (whose ``application [...] to the different physical phenomena is sufficient to teach us what we can reasonably hope to know about them.”): energy conservation, the principle of relativity (i.e.,\ invariance under uniform motion), the Second Law of Thermodynamics (``le principe de Carnot”), the principle of least action, Newton’s action-reaction principle and his Second Law, and mass conservation (``principe de Lavoisier”).  The principles derive their force from being bold generalisations of experimental findings (``résultats d'expériences forcement généralisés”) and from their broad scope (``généralité”) (p.\ 306). Conformity to those principles intimates the direction of the most promising theories and models that scientists should adopt and pursue. Conversely, these principles are, and ought to be, regarded as provisionally inviolable (``hors des atteintes de l'expérience”, p.\ 322): ``only after sincere efforts to save them must the principles be abandoned'' (p.\ 319, our translation).

Principles can, however, reach a stalemate in their heuristic power---their chief \textit{raison d’\^{e}tre}: once they cease to be fecund in significant problem contexts, experience ``condemns" them, without direct empirical contradiction (p.\ 323). This scenario, Poincaré declares, marks ``the current state of mathematical physics". The ``serious crisis” of Poincaré’s time consists in the repeated failure to account for several phenomena (viz.,\ the lines in emission spectra of atoms, null experiments of the ether, radioactivity, and the electrodynamics of moving charges), given the constraints of said principles: ``no matter how many hypotheses one piles up, one cannot satisfy all the principles at once; until now, one has only managed to safeguard some by sacrificing others” (p.\ 319, our translation). 

On Poincaré’s analysis, a crisis looms whenever principles that \textit{up to now have and ought to have} guided the development and appraisal of (empirically, in the main, underdetermined) theoretical options seem to have depleted their heuristic or problem-solving potential. Certain phenomena mandate that one either drop some of those principles, or replace them by new ones. The more drastic those requisite revisions---the graver the crisis.  

Poincaré's characterisation of a crisis spotlights the hurdles for a principle-based research strategy---a principled path forward.\footnote{Poincaré and Einstein weren't alone in stressing the role of principles in theory construction. As a \textit{distinctive style of doing physics}, it was a common theme in the late 19\textsuperscript{th} and  early 20\textsuperscript{th} century \parencite{Giovanelli2022MotivationalKantianism, Giovanelli2023RelativityPrinciples}.} His ``brief diagnosis" captures the sense in which the DM problem counts as a crisis. Furthermore, we also concur with both Einstein and Duhem (and, as we’ll see, a number of other luminaries in the history of physics) that principles play a crucial---but largely neglected,\footnote{For instance, one looks in vain for insights on principles in standard monographs on the scientific method (e.g.,\ \textcite{NolaSankey2007TheoriesScientificMethod}). The general role of principles in theory generation and justification is primarily discussed in the context of methodological reflections on quantum gravity (e.g.,\ \textcite{Holman2014FoundationsQG, Crowther2021DefiningCrisisQG}). It’s also acknowledged in the context of high-energy physics (\cite{Fischer2023NaturalnessPrinciples, Fischer2024NoLoseTheorems, Fischer2024PromiseOfSupersymmetry}). A first, more sustained recent recognition of principles in present philosophy of science---at a more abstract and descriptive level---is found in \textcite{DardashtiFischerHarlander2025_LifeCyclePrinciples}.} at least in contemporary philosophy of science discourse---role in theory/model construction in science. 

Despite our wholehearted agreement with the significance of principles, questions and substantive work remain: how best to construe `principles'? How exactly are they utilised in theory and model construction? How do they earn their privileged status, and how are principles justified more generally? These questions form the backbone of the present paper and the angle from which we’ll revisit the DM crisis.  The central thesis that we will defend here is that the DM crisis is best understood as the---at least temporary---exhaustion of a form of principle-guided inquiry: certain independently well-motivated principles have ceased to provide effective heuristic guidance for theory construction.

The \textbf{plan of the paper} is as follows. \textbf{\S\ref{sec:principles}} will outline a general account of principle-guided explorative research (`principled inquiry')---a proposal for understanding principles and their role in theory/model building. After reviewing the DM problem in \textbf{\S\ref{sec:DMparadigm}} and the principles therein in \textbf{\S\ref{sec:principlesofthedmparadigm}}, we'll consider in \textbf{\S\ref{sec:DMcrisis}} the specific principles which are either foregrounded or abandoned in each of four main approaches to the DM crisis. As adumbrated by Poincaré, the DM \textit{crisis} is, we’ll argue, the current predicament of DM research where guidance by principles has run into an impasse. We’ll summarise our main points and conclude in \textbf{\S\ref{sec:close}}.

\section{Principles and scientific theories}\label{sec:principles}

This section moots a functionalist conception  of principles, especially congenial to practice in theoretical physics, model and theory construction in particular (\textbf{\S\ref{ssec:functionalismprinciples}}). \textbf{\S\ref{ssec:principledinquiry}} examines more closely reliance on principles as a research strategy---`principled inquiry'---for pushing the envelope of what is known. In \textbf{§2.3} we briefly demarcate principled inquiry from a reactionary stance simpliciter, and from Popper's, Quine's, and Norton's methodological views. 

We'll harvest the fruits of our philosophical labours here in \textbf{§\ref{sec:principlesofthedmparadigm}} and \textbf{§5}. There, we'll demonstrate that principles as we conceive of them dovetail with the dominant DM research strategy, and that an analysis in terms of principles underwrites a novel perspective for nuanced critical evaluations of mainstream DM research. 

\subsection{Functionalism about principles}\label{ssec:functionalismprinciples}

Taking up a cue from \textcite{Stump2017},\footnote{And following a wider functionalist trend in recent philosophy of physics (e.g.,\ \textcite{Wallace2012, Read2018, Knox2013, Knox2019, WallaceKnox2023}).} we propound that principles be best understood via their---actual or intended---respective functions within a theory or model. That is, we submit that the status of a claim as a principle is \textit{defined} by the functions it serves or is employed for.\footnote{We thereby reverse the (more standard) way of thinking about principles, found in, e.g.,\ \textcite{Crowther2021DefiningCrisisQG}, or \textcite{Fischer2024NoLoseTheorems}, who discuss the functions that \textit{antecedently} defined principles fulfil.} 

The benefits of such functionalism are threefold. First, it grants primacy to scientific practice, i.e.\ to what scientists \textit{do} with principles---rather than starting from philosophical hunches. Secondly, functionalist approaches steer clear of strong---and therefore controversial---metaphysical or epistemological commitments (e.g., about the truth-conduciveness of induction or certain theoretical virtues). Thirdly, the relative philosophical neutrality affords a pluralism-friendliness, which we deem advantageous. It allows compatibility with more specific philosophical views (e.g., neo-Kantian or pragmatist approaches to the a priori, which have historically been bound up with reflections on principles, see, e.g.,\ \textcite{Stump2017}).

In this spirit, we stipulate that a principle be any claim that plays---or is adopted for the purpose of playing---one of two types of roles:\footnote{The terminology is deliberately echoing that common in neo-Kantian or Kant-inspired literature (especially \textcite{Poincare1902} and \textcite{Friedman2001, Friedman2002, Friedman2008}). The links between that tradition and ``principled inquiry'', i.e., our functionalist account of principles, are philosophically rich. Nothing in our discussion, however, hinges on overtly Kantian commitments. In particular, we disown any notion of apodictic knowledge and apriorism more generally, embracing a thoroughgoing fallibilism instead (cf.\ \cite{Stump2017}, esp.\ chs.\ 1 \& 9). It lies outside of the present paper’s ambit to spell out those links  (cf.,\ however, \textcite{HerfeldIvanova2021} and the contributions to their special issue). Instead our goal is to highlight how naturally principled inquiry meshes with scientific practice in innovative theorising.}   

\begin{itemize}
\item \textit{Regulative} principles guide research (without determining more specific content, cf.\ 
\textcite[p.\ 64]{Friedman2001}, \textcite{Giovanelli2022MotivationalKantianism}). They comprise claims that act (or are supposed to act) as ordering principles and programmatic directives, structuring and piloting inquiry. 
\item \textit{Constitutive} principles fulfil a more foundational function: as enabling conditions---or necessary premises---for theories and models (cf.\ \textcite{Darrigol2020}). We can parse them into three sub-types: 
\begin{enumerate}[(i)]
\item Conceptual-formal prerequisites for defining the theory/model’s entities or stating their relations and the laws they obey,
\item ``Coordinative principles” (\textcite{Reichenbach1920, Friedman2001}\footnote{\textcite{Padovani} has emphasised how \textcite{Reichenbach1920} has a rich view of constitutive principles that includes enabling conditions for experimental access. Although our account of principles is in no way wedded to the views of Reichenbach (or other historical neo-Kantian authors), this historical remark serves as a reminder that the envisioned coordination between formalism and empirical content isn't supposed to be flat-footed operationalism.}) that mediate (“coordinate”) between formal-mathematical structure and empirical reality, and in virtue of which mathematical equations acquire physical-empirical content,\footnote{These mustn’t be conflated with Reichenbach’s (\citeyear{Reichenbach1928}) ``coordinative \textit{definitions}”: albeit intended to serve the same function as what we call ``coordinative principles", Reichenbach's underlying operationalist semantics is widely rejected as na\"{i}ve.} and 
\item Adequacy constraints or selection rules that are imposed as (typically high-level) sine qua nons for viability.
\end{enumerate}
\end{itemize}

Regulative principles form a motley spectrum of injunctions. It spans adherence to abstract and malleable methodological precepts (e.g., testability and shunning of ad-hocness, or considerations of parsimony or simplicity), as well as quite ``tangible” meta-theoretical desiderata which can be squarely implemented as heuristic strategies (such as the amenability to the Least Action Principle \parencite{Stolzner2009}, or naturalness in high-energy physics (e.g.,\ \cite{Fischer2023NaturalnessPrinciples}), or UV completion in quantum gravity \parencite{CrowtherLinnemann2019}).  

The historical paradigm for constitutive elements in the sense of (i) is topological or geometric spacetime structure, as presupposed by mechanics and field theory (see, e.g.,\ \cite{Friedman1983}). Conceptually-formally constitutive elements aren't restricted to mathematical paraphernalia: for instance, the Zeroth and Minus First Laws for thermodynamics \parencite{BrownUffink}, or the conditions of the Born--Oppenheimer approximation for chemistry (which needs stable molecular structure) \parencite{HuggettLadyman} are examples of more physical/non-formal hue.\footnote{An important class of conceptually-formally constitutive elements of dynamical theories, such as hydrodynamics, electromagnetism or general relativity, are what \textcite{Curiel2017} calls ``kinematic constraints” (see \textcite{March2024} for further discussion).} 

Coordinative principles---constitutive principles in the sense of  (ii)---allow theories to be applied to the world. Here, we can't aspire to a more systematic answer to the deep question of how theories represent target systems (see, e.g.,\ \textcite{nguyen_frigg_2021_modelling, suarez_2024_inference} for recent magisterial analyses). We confine ourselves to observing that some assumptions are essential for such bridging a formalism and reality (cf.\ also \textcite{curiel_2019_schematizing, curiel_2020_on_propriety}). Historically, the postulate that time or length quantities be surveyed by rods and clocks (e.g.,\ \textcite{giovanelli_2014_but_one_must_not}; \textcite[§3.1]{Darrigol2020}) springs to mind. Also more domain-specific examples can be cited: the Boltzmann Principle (which links microstates and macrostates, thereby tethering microphysical descriptions to empirically accessible macrophysics), the Born Rule of quantum mechanics, or the standard ``interpretative principles" of GR, connecting its metric structure with the (idealised) behaviour of particles and light-rays (\cite[Ch.\ 2.1]{Malament2012}; cf.\ \cite{adlam_linnemann_read_2025_constructive}).

Our final sub-type (iii) of constitutive principles, adequacy constraints or selection rules, come in a variety of forms. They encompass more theoretical, even metaphysically flavoured, assumptions (such as causality conditions, or stability considerations), as well as more empirical ones (such as the Pauli exclusion principle for non-relativistic quantum mechanics, or conservation laws for energy or momentum). 

To cast the profile of the advocated functionalism about principles into sharper relief, some remarks are in order:
\begin{itemize}
    \item The categories of taxonomy of principles is neither mutually exclusive nor clear-cut; as is common to functionally individuated kinds, overlaps and cross-classifications exist. For instance, Lorentz invariance excels in both its regulative use (with tremendous heuristic power) and in its constitutive use (e.g., as the cornerstone of all of relativistic field theory, classical or quantum).
    \item We underscore the \textit{dynamical} and \textit{relativised} nature of the label `principle':\footnote{The terms are borrowed from \textcite{Reichenbach1920} and \textcite{Friedman2001}, respectively.} what counts---serves or is intended to serve---as a principle can change over time. As physics progresses, some claims may forfeit or gain the status of principles, concomitantly with their justification (invariably couched in our background knowledge---a topic to which we'll return shortly). For instance, until the advent of General Relativity conservation of energy and momentum were viewed as fundamental and universal principles; General Relativity revealed them to be contingent on special---and far from generic---symmetries of spacetime (see, e.g.,\ \textcite{Duerr2020}). In fact, principles can even be relinquished wholesale, as was the case, for instance, with the ergodicity hypothesis (the assumption that a system over time would eventually visit all the possible microscopic states). In Boltzmann's and Maxwell's pioneering work in Statistical Mechanics, it was a central principle. Later it was proven to be false for most realistic systems. Modern approaches accordingly dispense with it. 
    \item To regard a claim as a principle presupposes neither its truth nor epistemic certainty about it. Sometimes, we needn't even assign principles truth values at all. Consequently, for principle-hood it's not necessary that principles be \textit{empirically} well-confirmed. This might be impossible for practical reasons (think, for instance, of the Cosmological Principle---the assumed homogeneity and isotropy of the universe at large scales---before the discovery of the Cosmic Microwave Background in the 1960s; see, e.g.,\ \textcite{partridge_2019_cmb}), or because the inherently idealised or abstract nature of the principle in question stymies empirical tests (as is the case for Newton's First and Second Law). Instead, what matters for a principle's justification is that one possess compelling reason for deeming it functionally apt (more on this in \textbf{§\ref{ssec:principledinquiry}}).
    \item Finally, we stress that principles don't \textit{exhaust} the compositional elements of a model or theory. Due to their special role in the architecture of a theory or model, principles redound building blocks to them---but they don't make up the full theory or model. Principles must be supplemented by further creative leaps, theoretical or empirical input (for fixing parameters or specifying free functions), and/or boundary and initial conditions.
\end{itemize}

The plethora of examples given encourages the functionalist account of principles as a promising working hypothesis. In \textbf{\S\ref{sec:principlesofthedmparadigm}}, we’ll show its fertility in action. 

Let's next inspect the methodological peculiarities of the research strategy that leans on principles---what we'll dub ``principled inquiry". In particular, it's encumbent on us to address the question of evaluative standards for principles, criteria for \textit{good} principles. 

\subsection{Principled inquiry}\label{ssec:principledinquiry}

Principled inquiry is a research strategy that seeks to develop new theories or models by dint of principles. It proceeds not from scratch but by drawing on certain claims and ``elevating” them to principles (in the functionalist sense of \textbf{\S\ref{ssec:functionalismprinciples}}): one presupposes them in one’s research, harnessing them for regulative or constitutive purposes. In widespread scientists' parlance, the principles delimit the ``theory space” which is subsequently combed through when the theories or models so constructed are tested.\footnote{To pre-empt misunderstandings, it’s important to distinguish principled inquiry from two historical positions closely associated with the use and centrality of principles in physics. The first is the tradition of the `physics of principles' (see, e.g.,\ \textcite[ch.\ 2]{Giedymin1982ScienceConvention}). It denotes an approach that seeks to formulate increasingly general mathematical principles capturing the common empirical and structural content of rival physical theories. Variational principles are a paradigmatic example. Historically, this approach was often contrasted with a `physics of models', which aims to explain phenomena by constructing detailed representations of the underlying mechanisms, processes, and causal factors responsible for them. We wish not to perpetuate this dichotomy. Principled inquiry concerns the use of principles both in the construction of models and in more abstract forms of theorising. In particular, principles needn’t function as an \textit{alternative} to models; instead, they guide, constrain, and organise their construction. 

Also the distinction between principled inquiry and what \textcite{Einstein1919WhatIsRelativity} called `principle theories' deserves to be drawn. Here, we won’t join the fray over details of Einstein’s taxonomy, such as the appropriate ontological stances, epistemological questions or formal characterisations (see, however, e.g.,\ \textcite{Flores1999EinsteinTheories, Brown2005, VanCamp2011PrincipleConstructive, Frisch2011PrincipleConstructive, Lange2014PrincipleTheories, giovanelli_2014_but_one_must_not, Giovanelli}). Three key salient points stand out, however. First, as stressed in the foregoing point, principled inquiry is a research strategy for scientific innovation \textit{in statu nascendi}---an approach for \textit{constructing} novel theories or models. Einstein’s taxonomy, by contrast, is concerned with a classification of (\textit{finished}) theories (cf.,\ however, \textcite{Giovanelli}, who accentuates also this `logic of discovery' aspect of Einstein’s distinction). Secondly, Einstein presents a classification of types of theories, tracking different features, realist commitments, epistemological status and the kinds of explanatory achievements that are underwritten. Principled inquiry, by contrast, is decoupled from such a classificatory project. Thirdly and finally, for Einstein principles are empirical generalisations, whose justification lies in the supreme evidential support they enjoy. Principled inquiry, by contrast, allows for a more general and pluralist origins and rationales for principles.}

That metaphor insinuates a somewhat passive search within a well-defined intellectual territory stretched out before the scientist. This impression is misleading. Principled inquiry furnishes a powerful \textit{ars inveniendi}: a guideline or rule of thumb for generating scientific theories or models. To pre-empt misunderstandings we underscore two caveats of this heuristic method: in two regards principled inquiry is far from a mechanical algorithm nor a rigid and universal `discovery method' (cf.\ \textcite{Laudan1977ProgressProblems, Nickles1989HeuristicAppraisal, Nickles1990DiscoveryLogics, Nickles2005ProblemReduction}). First, unflinchingly fallibilist, principled inquiry doesn't automatically produce true theories or models, or even empirically adequate ones. More tentatively, together with skill and a modicum of luck, it allows scientists to construct prima facie \textit{promising} ones, worthy of further investigation. Whether  they eventually live up to our epistemic standards must be assessed in a second, later step (in what \textcite[p.\ 108]{Laudan1977ProgressProblems} has called the ``context of acceptance"). Secondly, principled inquiry captures a rough-and-ready ansatz---\textit{not} a fixed scheme with placeholders. To breathe life into it, background knowledge, sensitive to the context at hand, must be invoked: it supplies the principles that are apposite and the reasons in virtue of which we should regard them as apt for the cognitive purposes. Even with the principles chosen, principled inquiry doesn't prescribe \textit{how exactly} they should be assembled. Not even the existence of a model or theory that respects all the initially selected principles is guaranteed. 

How does a claim \textit{merit} selection as a principle? In line with our functionalist approach, we propose the following criterion, which pivots on the adequacy for a principle's purpose (cf.\ \textcite{Parker2020AdequacyPurpose}): principles qualify as methodologically sound when we have reasons for regarding them suitable for their intended function (regulative or constitutive) within the envisaged research context. We justify the adoption of a principle via plausible arguments that it conduces to the realisation of those functions. Such justifications depend on the scientific background knowledge available at the time. Not rarely are they intertwined with (explicit or tacit) convictions of more philosophical ilk.\footnote{For instance, Popperian falsificationism (inter alia) was expressly adduced by advocates of the Perfect Cosmological Principle in the context of Steady State Cosmology (\textcite{Kragh2022}; see also \textcite{Balashov1994} especially for the role of uniformitarianism). For cosmology more generally, \textcite[esp.\ \S8F]{ellis_2006_issues_philosophy_cosmology} underscores the need of a ``philosophical basis'' for principles  ``to shape the theory'' (op.\ cit.,\ p.\ 33), explicitly citing the underpinnings of the Copernican Principle.} The deliberative judgements that those reasons enter into allow for rational disagreement\footnote{As is common for judgements in science, \textcite{Brown2000}, and philosophical reasoning more generally, see, e.g.,\ \textcite{Elgin1996}.} and, consequently, for the pluralism witnessed and desirable in actual science, especially on the frontier of knowledge (see, e.g.,\ \textcite{Chang2012}). 

A few general things, however, can be said about justifying principles, at least \foreignlanguage{greek}{ὡς ἐπὶ τὸ πολύ}. Since regulative and constitutive principles perform different epistemic functions, standards appropriate for their justification likewise tend to differ:
\begin{itemize}
    \item The more abstract and general regulative principles are geared towards ensuring or conducing to widely shared methodological norms or scientific values, such as explanatory power, coherence, scope, or empirical accuracy. Whether in fact they do so is requires non-trivial argument. Many regulative principles operating as meta-theoretical desiderata can be traced back to such norms (together with physical background assumptions): the use of UV-completion in quantum gravity research (i.e.,\ the idea that a theory should formally hold up to all energy scales), for instance, is naturally viewed as a regulative admonition to maximise testability or predictive power (cf.\ \textcite[\S4.2]{Crowther2021DefiningCrisisQG}; \textcite{CrowtherLinnemann2019} for an exemplary \textit{critical} re-assessment of the principle's justification). 

    \item More specific regulative principles often encode experience-based rules of thumb (or strategies that have proven effective in the past and that are now, for better or worse, projected onto future research) or pragmatic considerations of cognitive costs or feasibility (e.g., in the form of Occamist appeals to simplicity and parsimony). For instance, in the context of Beyond Standard Model physics, the effective field theory framework, with its basic principle of separation of scales, is a case in point (see, e.g.,\ \textcite{KoberinskiSmeenk2023, KoberinskiSmeenk2023CosmoConst} for critical discussions in the context of Dark Energy and inflation, respectively).
\end{itemize}

Rationales for the three types of constitutive principles muster slightly different kinds of arguments:
\begin{itemize}
\item \textit{Conceptual-formal principles} are buttressed by a combination of---typically, a trade-off between (cf.\ \textcite{DuerrFischer2025})---fruitfulness (i.e.\ the resources for engendering innovation, expansion of cognitive horizons, formulating new questions and problems, and capacities for tackling them, see \textcite{Ivani2019, haufe_2024_fruitfulness}), and more pragmatic factors, related to feasibility, manageable cognitive costs, or qualitative understanding. The shining exemplar here are Newton’s Laws (see \textcite{Smith2014} for an authoritative study of their fruitfulness in celestial mechanics). 

\item \textit{Coordinative principles} should facilitate contact between theory and observable phenomena and testing in particular. Aptness for that function can be assessed along two chief axes: (i) realisability and robustness: good coordinative principles can be multiply (and consistently) realised via different setups; they also remain robust under variations in physical details of  such realisations and/or (small) modifications in the underlying theoretical assumptions; (ii) empirical richness and theoretical integration: good coordinative principles display numerous and diverse empirical connections; they should also cohere with, or be smoothly integrable into, our more theoretical background beliefs.

The geodesic principle---the fact that light-rays and point-particles trace out geodesics of the general-relativistic metric---exemplifies this. It admits of numerous realisations (including planets, satellites, atoms in free-fall interferometers, photons, gravitational waves or binary pulsars) and is smoothly integrated into General Relativity (where it comes close to becoming a theorem---see, e.g.\ \textcite{GerochEhlers, GerochJang1975, Dold-2025, puetzfeld_etal_2015_equations, GerochWeatherall2018}---and also holds for many modifications of GR).

\item For \textit{adequacy conditions}, empirically corroborated assumptions enjoy pride of place. For instance, the firm status of the Pauli exclusion principle as a constraint (long before it was proven in the context of the spin-statistics theorem, see \textcite{Massimi2005} for a comprehensive survey) rested on overwhelming empirical credentials: amongst others, atomic spectra and the anomalous Zeeman effect, the stability of atoms, or chemical valence.

\end{itemize}

None of these considerations guarantees that principled inquiry will succeed. Principles may turn out to be misguided, mutually incompatible, or simply unfruitful. Nevertheless, principled inquiry seems a prudent strategy for what one might call ``epistemic triage": the initial sorting and prioritisation of investigative paths---the first passes when (cognitive and material) resources are scarce and the first, and usually provisional, steps at exploration are undertaken. The more desperate the situation\footnote{From contemporary physics, the plights of Beyond Standard Model physics (e.g.,\ \cite{Hossenfelder2018, King2025}), cosmic inflation (e.g.,\ \cite{KoberinskiSmeenk2023CosmoConst}), or the Dark Energy problem (e.g.,\ \cite{wolf_duerr_2024_inflation}) spring to mind. In all three cases the standard principles haven't yielded the hoped-for success. Many scientists' patience has been exhausted. Pluralistic pursuit of also more radical ideas now indeed seems to be called for.} becomes---after principle-guided efforts have persistently not borne fruit---the more avenues should be investigated that jettison principles, and that make ever more daring forays into scientific speculation. 

Might the Dark Matter problem---frequently proclaimed one of the most urgent crises of contemporary physics---reached such a stage of exasperation?\footnote{For instance, \textcite[p.\ 52]{BertoneTait2018} enunciate: ``In light of this situation, the new guiding principle should be ‘no stone left unturned’: we should look for dark matter not only where theoretical prejudice dictates that we ‘must’, but wherever we can”.} Before embarking on a closer analysis of that question, some readers may wish to juxtapose principled inquiry and other research strategies for theory change. Other readers can skip this primarily philosophical intermezzo.  

\subsection{\textit{Intermezzo}: positioning principled inquiry}\label{sec:intermezzo}

Here, we'll contrast---via four queries---principled inquiry with other research strategies for theory change. 
\paragraph{\textbf{(Query 1)} \textit{Where do principled inquirers stand vis-à-vis Popper's methodology of `conjectures and refutations' \parencite{popper_2002_conjectures}?}} 

\textit{Reply:} They propose what they take to be a more realistic and effective alternative. For \textcite[p.\ 7]{popper_2002_logic}, ``(t)he initial stage, the act of conceiving or inventing a theory, seems [...] neither to call for logical analysis nor to be susceptible of it.” He relegates questions about the search for new theories to the ``psychology of knowledge” (ibid.). Principled inquiry is precisely concerned with that initial stage. It aspires to something akin to a (fallible and defeasible) ``logic of discovery"---whose existence Popper flat out denies (see also \textcite{Nickles1990DiscoveryLogics} for a critical assessment).

Popper declines to offer substantive \textit{prospective} advice about theory and model construction. The quest for scientific knowledge boils down to trial and error.\footnote{``The way in which knowledge progresses, and especially our scientific knowledge, is by unjustified (and unjustifiable) anticipations, by guesses, by tentative solutions to our problems, by conjectures. These conjectures are controlled by criticism; that is, by attempted refutations, which include severely critical tests. They may survive these tests; but they can never be positively justified" \parencite[p.\ xi]{popper_2002_logic}. Popper's denial of any forward-looking methodological advice is arguably the flipside of his anti-inductivism.} Scientists, according to Popper, do and ought to put forward ``audacious"---expressly \textit{im}probable (e.g.,\ op.cit.,\ p.\ 206)---guesses and otherwise, as the falsificationist's supreme methodological duty, subject ``these marvellously imaginative and bold conjectures or ‘anticipations’ of ours" (p.\ 278) to strict tests.

Both descriptively (a historical claim that we can't fully substantiate in the present paper\footnote{The analyses of \textcite{Nickles1978ProblemsConstraints, Nickles1981WhatIsProblem, Meheus1999Clausius}---very close in spirit to the present paper---show that problem situations in science are richly structured in terms of constraints, and that scientists systematically explore the space defined by these constraints, rather than make blind guesses.}), and normatively, principled inquirers repudiate such a trial-and-error methodology as inadequate. Exceptional circumstances---a crisis---however, may sway them to condone it as a last resort.

\paragraph{\textbf{(Query 2)} \textit{How inherently reactionary---biased in favour of conservatism---is principled inquiry?}} 

\textit{Reply:} In and of itself---not at all! Principled inquiry has arguably sparked some of the greatest revolutions in modern physics, including special relativity \parencite[ch.\ 4]{Brown2005}, general relativity (cf.\ \textcite{Janssen2013, Norton2020}), Bohr’s model of the atom (cf.\ \textcite{Kragh2012BohrQuantumAtom}), or Dirac’s theory of the electron \parencite{Kragh2022}.\footnote{It deserves to be stressed that the latter has been celebrated as ``(showing) the great superiority of principles over the previous empirical method'' (op.cit.,\ p.\ 57).} It doesn’t invariably or \textit{per se} privilege already accepted beliefs---or, for that matter, principles. The normative force of principled inquiry as a research strategy isn't inherently tied to conservatism: its rational warrant flows from the (context-dependent) justification of the principles. What matters is which principles we select, how we justify and use them---their adequacy for our cognitive purposes. Principled inquiry is as revolutionary or reactionary as the use scientists make of the principles. 

\paragraph{\textbf{(Query 3)} \textit{Doesn't principled inquiry merely rehash Quine’s `maxim of minimum mutilation'?}} 

\textit{Reply:} No. Quine avers that we should revise our beliefs---in particular, concerning updated theories vis-à-vis new or anomalous data---in a fashion that keeps the impact on our whole ``web of beliefs" as small as possible; residual latitude is whittled down by pragmatic considerations (simplicity in particular).\footnote{``Each man is given a scientific heritage plus a continuing barrage of sensory stimulation; and the considerations which guide him in warping his scientific heritage to fit his continuing sensory promptings are, where rational, pragmatic.” \parencite[p.\ 46]{quine_1951_two_dogmas}} 

Note two subtleties characteristic of Quine's position. First, said web comprises the totality of our beliefs---\textit{all} of science; Quine champions epistemological holism. Secondly, Quine reins in this holism by imputing to the web a structure of asymmetric relations: the ``entrenchment” of the web’s elements (i.e., the degree of their (logical and inferential) interconnectivity) renders some parts more ``peripheral”, and others more ``central”. Quine’s ``maxim of minimum mutilation” \parencite[p.\ 5]{Quine1970PhilosophyLogic} counsels conservatism with respect to this structured web (subject to the empiricist's categorical imperative that the web as a whole accord with experience). 

Interpreting the maxim also as a rule for prospective theory development (prioritising conservative research), the juxtaposition with Quine brings to the fore three salient differences with principled inquiry:

\begin{itemize}
\item Principled inquirers needn’t buy into Quine's holism. Yet, the latter contains a valuable insight that they agree with: our knowledge forms an organic system, in which some elements occupy distinguished positions---elements that ceteris paribus it seems prudent to build on and work with in order to advance science. Principled inquirers identify these elements as principles (defined functionally). 

\item While concurring with Quine on the intricate internal structure of our knowledge, principled inquirers oppose his characterisation of this structure in terms of entrenchment (see \textcite[p.\ 33]{Friedman2001} for details, see also \textcite{Friedman2002}). The interdependencies are done greater justice to, principled inquirers demur, by focusing on the functional role that certain claims do and plausibly can play (cf.\ \textcite{Darrigol2008} for a kindred perspective). 

\item More generally, principled inquirers baulk at Quine’s austere empiricism. Instead, they plump for a more permissive range of justifications of scientific principles.
\end{itemize}

\paragraph{\textbf{(Query 4)} \textit{How does principled inquiry differ from Norton's ‘material theory of induction’?}} 

Norton’s (\citeyear{norton_2003_material_theory_induction, norton_2005_little_survey_induction, norton_2010_no_universal_rules, norton_2014_material_dissolution, norton_2022_material_induction, norton_2025_large_scale_induction}) key idea is that inductive inferences are warranted \textit{directly} by local/domain-specific facts: it’s these facts in virtue of which ``inductive principles” obtain—not in virtue of the principles’ formal schemes. Such inductive principles, i.e. factually supported assumptions, can be utilised in (ampliative) inferences. 

Insofar as their presence in theory and model construction is concerned (rather than the retrospective assessment of complete theories or models against available evidence---the primary focus of Norton’s discussion), we take Norton to suggest that such principles should either be preserved in new theories/models as constraints (at least in some limit), or figure more actively in the discovery of such theories/models via eliminative inductions \parencite{norton_1995_eliminative_induction}. In contrast to the material theory, principled inquiry doesn’t limit such justification to factual-empirical grounding (nor, consequently, limit principles to materially well-supported hypotheses).\footnote{Principled inquiry \textit{subsumes}, we surmise, most compelling cases of material inductive principles. For one of Norton’s favourite examples, `Ha\"{u}y’s Principle' from crystallography, it's straightforward to glean this.} Regulative principles don't seem to find a natural home in Norton's theory at all.

Apart from this difference, the comparison between principled inquiry and Norton’s ``material-inductive” approach highlights an important commonality: both dismiss the idea of warranting inferences by a \textit{universal, formal-logical} template. Instead, on both views, principles receive local and domain-specific justifications: they hinge on background knowledge about a domain---fallible and historically evolving.

\section{The Dark Matter problem}\label{sec:DMparadigm}

This section sketches the Dark Matter (DM) problem. We highlight the pickle that the DM problem lands researchers in with respect to the observational indistinguishability of the various DM proposals. As indicated in \textbf{\S\ref{sec:intro}}, however, such evidential underdetermination isn’t the whole story---a claim we’ll make good on in later sections.

Our present\footnote{More \textit{historically} oriented treatments are \textcite{Sanders2010} and \textcite{BertoneHooper2018}. Those authors seek a reconstruction of the DM problem from a \textit{present day} practising physicists’ understanding---occasionally somewhat at the expense of historical faithfulness. \textcite{DeSwart2020Closing, DeSwart2024FiveDecades, DeSwartEtAl2017} offer a more careful historical account proper.} understanding of the DM problem breaks into two parts:\footnote{\textcite{Allzen2024} rightly warns readers of the historical material against inherently linking the two parts, seductive as this may appear with \textit{our} benefit of hindsight. The historical continuity between non-luminous matter and our present-day notion of DM is more tenuous than often suggested.} an estimate of the amount of non-luminous matter, and the realisation that this matter isn’t baryonic (i.e., unlike ordinary matter, not composed of quarks), respectively.

The establishment of the now-consensus opinion that preponderantly invisible matter populates the universe is buttressed by sundry milestones:\footnote{We loosely follow \textcite[ch.\ 11.2]{LongairSmeenk2019} and \textcite{Roos2010}, also for further details.}
\begin{itemize}  
\item \emph{Local galaxy dynamics:} Pioneering work by Oort in the early 30s on stellar motions within the Milky Way showed that the total mass density in the solar neighbourhood was roughly twice that of the known stars: the dynamical mass---the mass inferred from the observed dynamical effects---significantly exceeded the visible mass, implying more invisible (`dark') mass than visible stars in our vicinity.
\item \emph{Galaxy cluster kinematics:} Likewise in the 1930s, Zwicky investigated the kinematics of galaxy clusters (primarily the Coma cluster), on the basis of virial considerations (that is, assuming a kind of statistical---but \textit{not} thermodynamic!---equilibrium). Their velocities turned out to be far too large to be held together by the luminous mass alone. The clusters, Zwicky concluded, contain about 10 times more mass than seen. 
\item \emph{Galaxy rotation curves:} Systematic and high-quality observations of spiral galaxies by Rubin and colleagues during the early 1970s revealed `flat' rotation curves: contrary to expectations, based on the visible masses (and Newtonian gravity), the stars' orbital speeds remain high---roughly constant in fact---far beyond the visible stellar disk. This suggests extra mass in extended halos. Subsequent measurements throughout the 70s and early 80s confirmed that this is a common pattern in spiral galaxies.
\item \emph{Disk stability arguments:} Beginning in the 1970s, simulations and theoretical considerations showed that, if matter were limited to luminous matter, thin disk galaxies would become dynamically unstable; on very short timescales they would form bars---in tension with the many thin disk galaxies actually observed. Embedding disks in a large, low-luminosity dark halo naturally stabilises them (\textcite{Ostriker}). 
\item \emph{Hot gas in clusters:} X-ray imaging of clusters since the 1980s disclosed that they are enveloped by hot, diffuse gas. The gas temperature and distribution imply total cluster masses far greater than the mass of galaxies alone. High-precision X-ray missions in the early 2000s have consistently shown that 80--90\% of cluster mass must be dark.
\item \emph{Gravitational lensing}: Under suitable conditions, galaxy clusters act as gravitational lenses. They can then directly map the projected mass distributions. As was demonstrated in 2006, the ``Bullet Cluster”---a configuration of two colliding clusters---exhibits a clear separation between the X-ray--bright gas and the lensing-inferred mass peaks: in contrast to the former, the Cluster’s dominant mass behaves collisionlessly, permitting largely unhindered interpenetration. Moreover, the \textit{total} gravitating mass inferred from the lensing effect (via General Relativity) is centred on the galaxies, not the gas; the former outweighs the latter, thanks to (according to the mainstream view) a sizable DM fraction. Other clusters have yielded similar conclusions.  
\item \emph{Large-scale structure surveys:} Galaxy redshift surveys between the late 1990s and the late 2010s, and weak lensing maps, reveal the cosmic web of matter across the Universe. The observed clustering of galaxies, clusters, and voids, and its evolution over time, matches simulations of a cold DM-filled universe. The standard $\Lambda$CDM model---with its matter and energy budget involving $\sim 5\%$ baryons, $\sim 25\%$ cold dark matter, and $\sim 70\%$ dark energy (effectively described by a cosmological constant $\Lambda$)---accurately reproduces the large-scale distribution of matter.
\item \emph{CMB anisotropies:} Beginning with the COBE satellite in 1992 and followed by the WMAP mission (2001--2010) and the Planck mission (2013--2018), minuscule temperature fluctuations in the cosmic microwave background (CMB) were measured with high precision, providing a snapshot of the Universe prior to the formation of neutral atoms (``recombination", ca.\ 380,000 years after the Big Bang). These fluctuations arise from primordial density and velocity perturbations in the tightly coupled photon–baryon fluid evolving within gravitational potential wells. When analysed via their angular power spectrum, the resulting acoustic peak structure encodes detailed information about the cosmic matter content. The relative peak heights---most notably the substantial amplitude of the third peak---require a large amount of gravitating matter that doesn’t participate in the photon–baryon oscillations. This excess gravitating component deepens potential wells before recombination, enhancing higher-order acoustic peaks. The inferred total matter density exceeds that associated with luminous matter alone by more than a factor of five.
\end{itemize}

These independent lines of evidence from sub-galactic to cosmological scales converge on the same conclusion: most of the gravitating mass in the Universe is dark. 

Now to the other lines of evidence that divulge something about its nature---albeit \textit{ex negativo}: unlike ordinary atomic matter (composed of baryons, which in turn are built up from quarks), DM isn’t baryonic:  

\begin{itemize}
\item \emph{Big Bang nucleosynthesis (BBN):} The abundances of light elements---most notably deuterium, helium-3, helium-4, and lithium---produced during the first few minutes after the Big Bang are exquisitely sensitive to the cosmic \textit{baryon} density. Standard BBN calculations (performed, in mature form, since the 1960s, see, e.g.,\ \textcite{Turner2022BBN}), when confronted with observed primordial abundances (especially deuterium), tightly constrain the fraction of baryonic matter. By roughly a factor of five, the resulting baryon density falls short of the total matter density inferred from dynamical and cosmological observations (i.e.\ the evidence rehearsed in the above first part of the DM case). Any additional gravitating matter must therefore be non-baryonic. The impressive accuracy of measurements of primordial deuterium in the late 1990s (and subsequent confirmation by CMB measurements in the early 2000s) renders BBN a ``linchpin in the case for non-baryonic dark matter” (op.cit.,\ p.\ 15).

\item \emph{CMB constraints on baryons:} The already-mentioned CMB anisotropy measurements that support the existence of DM also fix the baryon density. Baryons affect the oscillations of the photon–baryon plasma by enhancing its compression relative to its rarefaction. The effect can be gleaned from the ratio of peaks in the CMB power spectrum. Precision fits to the CMB power spectrum yield a baryon density consistent with BBN---and again far too small to account for the total matter density required by the arguments considered above. The excess matter mustn’t couple to radiation in the way baryons do.

\item \emph{Structure formation timescales:} In a purely baryonic universe, density perturbations cannot begin to grow efficiently until after recombination, when baryons decouple from photons. Numerical simulations show that this leaves insufficient time to form the observed large-scale structure by the present epoch. Non-baryonic matter, by contrast, decouples much earlier. It can begin clustering well before recombination, seeding the gravitational potential wells into which baryons later fall. This early growth is essential to reproducing the observed hierarchy of cosmic structures.

\end{itemize}

\textcite[p.\ 5]{BalaczEtAl2024} summarise the situation: ``(t)he plethora of cosmological observations cursorily mentioned above do not only precisely tell us how much dark matter there is in total, they also put stringent constraints on its properties. [...] Both BBN and CMB considerations also leave very little room for any form of energy conversion between these components, i.e.\ any mechanism that would dump energy from dark matter into the primordial plasma of ordinary matter [...]. Remarkably, cosmological considerations even put tight constraints on dark matter models that do not interact with ordinary matter in any appreciable way. [...] Furthermore, dark matter moving at slightly too large velocities after CMB times---coined `warm’ dark matter---would escape from overdense regions before they fully collapse, preventing the formation of small gravitationally bound objects such as dwarf galaxies. [...] Notably, these constraints also apply to dark matter models that do not involve new elementary particles.” 

The observational case for a discrepancy between the \textit{observed} luminous masses and the \textit{theoretically inferred} ones (i.e. the reconstruction of what mass ought to exist, given our background knowledge) is overwhelming. The cogency of that discrepancy jars with the astonishing multitude of hypothetical possibilities for resolving it.
The most popular---and still in-principle viable---options include (see, e.g.,\ \textcite{Bertone2010ParticleDarkMatter, BozorgniaEtAl2025DarkMatterCandidates} for more comprehensive discussions):

\begin{itemize}
\item \textit{Weakly Interacting Massive Particles (WIMPs)} are heavy, and predominantly interact weakly with ordinary matter (besides gravitationally). They arise naturally in extensions of the Standard Model and are produced thermally in the early Universe, before freezing out as cosmic expansion diluted them.
\item \textit{Axions} are extremely light particles whose coherent cosmic oscillations could behave like DM.
\item \textit{Fuzzy Dark Matter} consists of bosons with ultra-light mass. Their wave-like behaviour becomes apparent on galactic scales. Their enormous de Broglie wavelengths generate an effective quantum pressure that suppresses structure on the smallest scales, while preserving the successful large-scale predictions of the standard cosmological model.
\item \textit{Sterile neutrinos} are hypothetical neutrinos that interact only through gravity (and, possibly, partake of slight mixing with the known neutrinos). 
\item \textit{Kaluza--Klein particles} arise in theories with compact extra spatial dimensions, where familiar particles possess heavier excitations corresponding to motion in the extra dimensions. 
\item \textit{Primordial Black Holes (PBHs)} formed in the very early universe from unusually large density fluctuations. If produced in the appropriate abundance and mass range, they could constitute some or all of the DM.
\item \textit{Modified Gravity} proposals dispense with DM altogether. Instead, they modify the laws of gravity so that the observed motions of stars and galaxies can be explained without invoking additional unseen mass. 
\end{itemize}
These candidates span almost fifty orders of magnitude in mass---from fuzzy DM with masses around $10^{-22}\text{eV}$ to super-heavy relics near the Grand Unified or Planck scales ($10^{25}$--$10^{28} \text{eV}$).\footnote{Surveying the whole spectrum of candidates, \textcite[p.\ 51]{BertoneTait2018} even mention 90(!) orders of magnitude!} They differ dramatically in their physical motivation, mechanisms, cosmological histories, and avenues for detection. 

It’s all the more baffling and frustrating that at present we can’t adjudicate amongst them on the basis of evidence. Concerted efforts at experimental searches (through direct, indirect and terrestrial/collider-based, see, e.g.,\ \textcite[chs.\ 5--7]{BauerPlehn2018}) have hitherto been thwarted (see \textcite{Antoniou2023} for an instructive philosophical analysis).  
The status quo of DM research thus presents us with a catch-22. On the one hand, we have  compelling evidence for what \textcite{MartensLehmkuhl2020} call `dark phenomena'. On the other hand, our evidential predicament is ``unsatisfactory, since it leaves us with no clear idea of what dark matter actually is'' \parencite[p.\ 8]{BalaczEtAl2024}. What we \textit{know} are constraints that concrete (``fundamental'', \cite{DeBaerdemaeker2025}) DM models must respect. This `common conceptual core' is too `thin' \parencite{Martens2022Realism, Vaynberg2024}---``[resembling] in some relevant ways that of the early concept of genes” \parencite[p.\ 15]{Martens2022Realism}: the evidential underdetermination undercuts more specific commitments.

But what makes the DM problem so nerve-wracking transcends underdetermination. Lest they drown in this flood of DM candidates, scientists need guidance: which ideas might it be reasonable to prioritise, at least for the time being? It’s here that principled inquiry comes into its own. It articulates, as the subsequent sections will argue, the strategy and rationale that underlies the mainstream DM research---those DM proposals that have received most attention and research efforts: WIMPs, sterile neutrinos, axions, and primordial black holes. Moreover, it also explains the ``growing sense of ‘crisis’ in the dark-matter particle community, which arises from the absence of evidence for the most popular candidates for dark-matter particles—such as weakly interacting massive particles, axions and sterile neutrinos---despite the enormous effort that has gone into searching for these particles” \parencite[p.\ 51]{BertoneTait2018}: DM research seems to have reached the heuristic limits of principled inquiry---suggesting that this cautious standard research strategy might have to give way to something more radical.

Before adverting to the details of the key DM candidates (\textbf{\S\ref{sec:DMcrisis}}), we'll next clarify what mainstream DM research has in common. Vis-à-vis the heterogeneity just surveyed, the answer at first blush appears to be: very little. Principled inquiry, however, unveils a surprising degree of coherence.

\section{Principles of the Dark Matter paradigm}\label{sec:principlesofthedmparadigm}

In what follows, ``DM paradigm'' shall denote the theoretical and methodological framework within which mainstream DM research operates (cf.\ \textcite[p.\ 45]{Friedman2001}). Correlatively with the attention they today enjoy, we'll for the purposes of this article equate it with research on WIMPs, axions, (sterile) neutrinos, and primordial black holes---the subject of \textbf{§\ref{sec:DMcrisis}}. Building on our functionalist account of \textbf{§\ref{sec:principles}}, this section will flesh this out in terms of the pertinent principles: the constitutive ones, circumscribing the theoretical horizon of physics assumptions within which advocates of the DM paradigm strive to understand DM, in \textbf{\S\ref{ssec:constitutive-principles}}, and the regulative ones, delineating the paradigm's epistemic aims and methodological orientation, in \textbf{\S\ref{ssec:regulative-principles}}.  

\subsection{Constitutive principles of the DM paradigm}\label{ssec:constitutive-principles}

The DM paradigm's constitutive principles fall into two groups: those dealing with gravity and tied to General Relativity (\textbf{\S\ref{ssec:gravity-principles}}), and those dealing with non-gravitational physics and tied to the Standard Model of Particle Physics (\textbf{\S\ref{ssec:particle-principles}}).

\subsubsection{Principles of gravity: the general-relativistic framework}\label{ssec:gravity-principles}

To accommodate DM phenomenology, one chiefly has two options: (DM) the---predominant---Dark Matter paradigm, which keeps GR and posits suitable (usually: novel) types of matter, or (MOD-GRAV) the---minority---Modified Gravity approach, which trades in deviations from GR (typically hoping to get by with only standard matter).\footnote{\label{fn-28}We set aside here ``mixed" approaches that abandon GR \textit{and} posit new types of matter. From the perspective of principled inquiry, they inherit ``the worst of both worlds".}

In favour of (MOD-GRAV), occasionally (e.g.,\ \textcite{LahavMassimi2014DarkEnergy, Merritt2020PhilosophicalMOND}) what at first blush seems like a historical analogy is invoked. It asserts that, just like DM phenomena today, Mercury's perihelion precession used to be an anomaly that defied a satisfactory explanation within Newtonian gravity, and that called for the latter's abandonment. That explanation eventually arrived when GR superseded Newtonian gravity.

It's instructive to push back against the allure of this analogy, and, correlatively, against the strategy (MOD-GRAV), from the perspective of principles. The modifications in question, i.e.\ the general-relativistic correction terms to the potential of the Kepler problem, merely alter the specific form of the gravitational law---a \textit{particular} interaction. They don't affect the validity of the \textit{more general} principles of Newtonian physics (viz.,\ Newton's axioms with their Euclidean spacetime structure and kinematics, alongside conservation laws for energy, momentum and mass). 

GR, by contrast, can't be tampered with so easily, without sacrificing core principles. Following \textcite{Lovelock1971EinsteinTensor, Lovelock1972FourDimensionality} (cf.\ \textcite[§2.4.1]{CliftonEtAl2012ModifiedGravity}), let's review a natural set of axioms that leads to GR, and briefly indicate the rationale of their underlying principles:\footnote{Alternative paths, based on different sets of principles, are possible (most prominently the spin-2 approach, based on principles of gauge theory, see, e.g.,\ \textcite{Feynman1963QuantumTheoryGravitation, Deser1970SelfInteraction}). We choose the most standard approach, whose principles (to our minds) are especially compelling.} 

\paragraph{Gravitational effects are represented by the metric tensor $g_{\mu \nu}$ (and its derivatives) of a four-dimensional pseudo-Riemannian spacetime geometry.} 
    
What underlies this is the Weak Equivalence Principle, which is essentially equivalent to the equality of inertial and gravitational mass of test particles (see, e.g.,\ \textcite{Norton1985EquivalencePrinciple, DiCasolaEtAl2015Nonequivalence, Lehmkuhl2022EquivalencePrinciples, ReadTeh} for details). Empirically, it's a robust finding, serving a constitutive function for GR \parencite{Friedman2001, Friedman2009EinsteinKantApriori}. Thanks to the (logically stronger) Einstein Equivalence Principle, likewise well-confirmed (at least at the scales of the solar system), gravity and inertia are even more intimately entangled: they can be represented in a unified way by $g_{\mu \nu}$ (cf.,\ for instance, \textcite{Lehmkuhl2014GeometrizesGravity}). In tandem with this appealing feature, the unification also contains a pragmatic dimension, pivoting on parsimony: gravitational effects are encoded in one single object (with no additional fields). Finally, the choice of the geometric principles of Riemannian geometry as the formal-conceptual arena for GR is likewise warranted by simplicity considerations. That is, from the perspective of more general geometries, Riemannian geometry counts as arguably the simplest generic extension of Euclidean geometry.

\paragraph{Gravity couples universally and minimally to matter.} By this is meant that (i) we obtain the general-relativistic matter Lagrangian $L_M$ from the special-relativistic one via the ``substitution rules'' \parencite[§3.4]{Wald}

\begin{equation}
\left\{
\begin{array}{c}
\eta_{\mu\nu} \\
\partial_\mu
\end{array}
\right.
\longrightarrow\;
\left\{
\begin{array}{c}
g_{\mu\nu} \\
\nabla_\mu
\end{array}
\right. ,
\end{equation}
(ii) this prescription holds for all types of matter, and (iii) the total Lagrangian (whose variation with respect to $g_{\mu\nu}$ and the matter fields yields the full field equations) is given by $L_{\text{tot}} = L_{\text{grav}} +\kappa L_M$ for some (still to be determined, as per the Newtonian limit, see below) coupling constant $\kappa$. 

The coupling scheme rests largely on the Einstein Equivalence Principle, as mentioned already above (alongside considerations of simplicity).  

\paragraph{Field equations for $g_{\mu\nu}$ should at most be of second order.} 
    
This regulative principle is at bottom a pragmatic desideratum, adopted for reasons of (mathematical) simplicity (and inspired by the analogy with electromagnetism). Higher-order derivatives tremendously complicate the dynamics of a theory. In fact, they can engender conceptual pathologies, such as instabilities.
 To eschew higher-order derivatives thus lowers cognitive costs, and hedges risks (by avoiding difficulties that vitiate viability).

\paragraph{Newtonian limit: For static, weak fields Newtonian gravity is recovered.} 
    
This ``correspondence principle'' fixes the free parameters. It also serves two other purposes. By reducing to the Newtonian limit, the empirical content and explanatory successes of GR's precursor are preserved. Thanks to its entrenchment in our background knowledge, that content acts as a viability constraint on any gravitational theory (or, in our above terminology: as an adequacy condition or selection rule). Moreover, the Newtonian limit also acts as a coordinating principle, bridging mathematics and reality: it anchors GR's formal structure in an empirical domain of application; from the latter physical and epistemic content and significance flow into the equations.

\bigskip

\noindent In sum, these principles seem persuasively motivated: a decision to adhere to them in tackling the DM problem would stand on firm rational ground.\footnote{The case for the DM paradigm is significantly strengthened further in light of the difficulties that (MOD-GRAV) incurs as a solution to the DM problem: empirical adequacy (uncertainty about which still persists) necessitates \textit{drastic} departures from widely cherished principles, such as introducing a preferred foliation of spacetime (see, e.g.,\ \textcite{Merritt2020PhilosophicalMOND, DuerrWolf2023_MethodologicalMOND}).} Principled inquiry into the DM problem, then, suggests that one keep GR, and---ideally likewise guided by principles---search for a suitable realisation of GR's energy-momentum source. 
Turning to non-gravitational physics, we'll next inspect such principles. 

\subsubsection{Principles of matter: the Standard Model of Particle Physics}\label{ssec:particle-principles}

The DM paradigm fully endorses the principles encoding ``gravitational orthodoxy" (\textbf{\S\ref{ssec:gravity-principles}}). It does require, though, \textit{some} departures from orthodoxy as regards matter (\textbf{§\ref{sec:DMcrisis}}). The Standard Model of Particle Physics (SM)---enshrining the established physics of matter---provides the starting point for DM candidates. As \textbf{\S\ref{ssec:regulative-principles}} will spell out, the guiding \textit{regulative} principle within the DM paradigm is conservatism or ``minimal mutilation'' with respect to the physics of matter (cf.\ Query 3 in \textbf{\S\ref{sec:intermezzo}}).

Our task here is to examine \textit{what} the DM paradigm bids us conserve as much as possible: the SM’s main principles and their various functions within the SM.
\paragraph{\textit{Quantum field theory (QFT).}} This provides the SM’s conceptual-formal architecture.\footnote{We’ll steer clear of details regarding the \textit{interpretation or ontology} of QFT. Instead, we are concerned with the working posits typically employed in scientific practice in applications and further model-building, keeping ontological and interpretative claims at a minimum.} Here, although strictly speaking encompassing several more specific principles, we treat QFT as an  ``umbrella principle'' to highlight this constitutive function, as per our functionalist account. QFT delimits the framework of countenanced physical entities and interactions: quantum fields whose interactions arise from local operators in a Lagrangian, and whose excitations correspond to particles (in a suitable ``emergent” sense, see \textcite{Falkenburg2007ParticleMetaphysics}; cf.\ \textcite{Fraser2022ParticlesQFT} for a more critical view). 

The rationale for QFT was that it appeared to be the minimal theoretical language capable of reconciling quantum mechanics with relativistic field physics, preserving locality and enabling the systematic description of particle creation, annihilation, and interactions. Historically, it emerged from extending---conservatively extrapolating---the lessons of relativistic wave equations (especially Dirac's theory of the electron), quantised radiation fields, and scattering theory. The spectacular successes of quantum electrodynamics consolidated the fertility of QFT as a framework. The SM’s other key theoretical pillars---quantum chromodynamics and the electroweak theory---were indeed explicitly modelled on quantum electrodynamics.

The field dynamics of QFT inherits the Born Rule from ordinary quantum mechanics. Also within the SM it functions as a coordinative principle: via QFT, one computes amplitudes or correlation functions for physical processes (such as scattering and decay), which are then converted into measurable probabilities and cross sections, thereby linking up the formalism and observable phenomena. 

\paragraph{\textit{Lorentz Invariance}.} The relativistic character of QFT brings with it a specific structural presupposition: conformity to the spacetime symmetries of SR. The QFTs considered in the SM must respect Lorentz invariance. This imposes a physically and heuristically substantive structural constraint (cf.\ \textcite[ch.\ 2]{WheelerNDRelativisticFieldTheory}): matter fields must transform according to appropriate representations of the Lorentz group, while particle states are classified by representations of the Poincaré group (see, e.g., \textcite[ch.\ 2]{Maggiore2005ModernQFT}). Lorentz invariance thereby constrains the possible types of matter and interactions independently of the particular internal gauge structure one may wish to adopt.

 \paragraph{\textit{Chiral electroweak structure.}} Only so-called left-handed fermions---one of two possible ``handed” versions of electrons, neutrinos, and quarks---transform non-trivially under the symmetry group ($SU (2)_L$, see below), distinctive of the electroweak interaction. The weak interaction acts on left-handed components of fermions, but not on their right-handed counterparts. Unlike the other fundamental interactions, the weak interaction isn’t invariant under mirror reflection (`parity'): nature distinguishes between left and right. The SM bakes handedness into its foundations: left- and right-handed versions of the same particle (defined mathematically via unique decompositions of a quantum field) are assigned different symmetry properties. Left-handed states are grouped into pairs (`doublets') that transform non-trivially under the weak symmetry, whereas right-handed states are treated as singlets that do not. 

Chirality functions as an adequacy condition (in our third sense of a constitutive principle) for the SM’s construction. Its justification stems from the surprising \textit{empirical} discovery of parity non-conservation. The \citeyear{WuEtAl1957Parity} Wu experiment showed that beta decay depends on the orientation of spin relative to emission direction, revealing a physically grounded preference for a particular spin--momentum correlation. Whereas the electromagnetic and strong interactions respect it, the weak force violates the above-mentioned mirror reflection symmetry. 
\paragraph{\textit{Gauge structure---the principle of local gauge invariance.}} The SM assumes that interactions are governed by laws invariant under continuous, local (point-dependent) internal symmetry transformations (transformations in additional `internal' spaces, attached to spacetime points). Matter fields are assigned conserved quantities (`charges') with respect to internal symmetry groups; the field variables transform according to representations (again in the technical sense) of that internal symmetry. Interactions are mediated by gauge fields introduced to preserve local symmetry. These interactions and their coupling heuristically arise naturally from the requirement of local gauge invariance. The dynamics itself is constructed from gauge-covariant quantities, while the quanta of the gauge fields correspond to the interaction carriers (see, e.g.,\ \textcite{Lyre2009GaugeSymmetry} for details).

As a constitutive principle in our first sense, gauge structure supplies part of the SM's conceptual-formal architecture.\footnote{Formally, the SM’s gauge-theoretic framework later found its mathematically most perspicuous expression in the language of fibre bundles: matter fields are represented as sections of bundles associated with a principal gauge bundle, while gauge fields correspond to connections on that bundle and their dynamics are encoded in the associated curvature.} Historically, the SM’s gauge-theoretic underpinnings were motivated by a generalisation of a structural lesson drawn from electromagnetism (as per Weyl’s reformulation thereof) and quantum electrodynamics: namely, that interactions may be systematically associated with local internal symmetries. In \citeyear{Yang1954}, Yang and Mills extended the gauge framework from the Abelian symmetry underlying electromagnetism to non-Abelian symmetry groups. Thereby, they inaugurated the template later realised in the SM. By the time quantum chromodynamics was formulated in the early 1970s, local gauge symmetry had already demonstrated its fertility in the electroweak sector. This made a gauge-theoretic treatment of the strong interaction a heuristically plausible principle to impose in the construction of quantum chromodynamics, the SM’s capstone.

The SM’s gauge structure doesn’t in and of itself uniquely prescribe the gauge \textit{group} (characterising the symmetries of the afore-mentioned internal spaces). In the SM, the internal symmetry is given by the direct product group $SU(3)_C \times SU(2)_L \times U(1)_Y$. The product structure indicates that these symmetry factors act on distinct sectors of the theory---rather than being unified in a more substantive sense into a single simple gauge group (cf.\ \textcite{Maudlin1996UnificationPhysics, Lehmkuhl2009SpacetimeMatters}). The subscripts indicate their physical roles: $SU(3)_C$ is the `colour' symmetry governing the strong interaction of quarks and gluons in quantum chromodynamics (where `colour' denotes a three-valued charge carried by quarks (analogous, in an abstract sense, to electric charge); $SU(2)_L$ acts on left-handed fermion doublets and encodes the chiral weak isospin structure of the electroweak interaction (where `isospin' is an internal quantum number that determines how particles participate in weak processes); and $U(1)_Y$ is associated with weak hypercharge, another internal charge. Together with weak isospin (after electroweak symmetry breaking) it yields the observed electromagnetic charge.

Rather than a \textit{deep}, fundamental principle (on a par with, say, Lorentz invariance), we deem the SM's gauge group $SU(3)_C \times SU(2)_L \times U(1)_Y$ a more contingent specification within the SM's gauge structure. Its justification is comparatively weak and pragmatic: this group turns out to be merely the most economical symmetry group structure, compatible with a number of highly non-trivial theoretical (e.g., anomaly cancellation as a consistency requirement) and empirical constraints (e.g., confinement phenomenology from hadron spectroscopy).\footnote{As targets of desiderata, these constraints likewise operate as principles, notwithstanding their fairly (context-)specific nature. This illustrates a point made earlier: whereas principle-hood is defined functionally, what counts as a principle lies on a continuum of generality (with the Least Action Principle as an exemplar, exhibiting exceptionally high generality). In this regard our functionalist account of principles differs markedly from Cassirer or Planck, who extol the generality of principles as their \textit{essential} feature \parencite{Giovanelli2023RelativityPrinciples, Giovanelli2024PracticePrinciples}.}  
\paragraph{\textit{Renormalisability.}} It demands that the class of admissible interaction terms in the Lagrangian (as per the insistence on QFTs) be restricted to those yielding a QFT whose ultraviolet divergences in perturbative calculations can be systematically absorbed into a finite number of physical parameters.

In the SM's genesis, renormalisability functioned as an adequacy constraint or selection rule: it was believed essential for mathematical consistency (avoidance or at least tractability of divergences) and predictive control over quantum corrections, excluding interaction terms that would require an infinite proliferation of counterterms.\footnote{E.g., the electroweak theory only became widely accepted after `t Hooft and Veltman demonstrated, in \citeyear{tHooft1972}, the renormalisability of spontaneously broken non-Abelian gauge theories.} 

Historically, renormalisability was invoked to pare down the space of viable QFTs for the development of the SM. Here another feature of our functionalist account of principles comes to the fore: changes in background knowledge can alter the status of an assumption as a principle.\footnote{Friedman’s apt term of the `\textit{dynamical} a priori' is intended to capture precisely this contingent, historically changing and hence revisable nature of principle-hood (see also \textcite[ch.\ 5]{Stump2017} for similar conceptions of the a priori in the early Reichenbach, C.I.\ Lewis, Dewey or Pap).} In the more contemporary framework of Effective Field Theory, developed from the 1970s onwards, renormalisability is no longer regarded as a sine qua non. Instead, it was recognised that one can preserve predictive power even in the face of non-renormalisability. Renormalisability was thus demoted to a more contingent desideratum of low-energy effective descriptions within a broader hierarchy of theories.

\subsection{Regulative principles of the DM paradigm}\label{ssec:regulative-principles}

Complementing the physically substantive constitutive principles, the DM paradigm also comprises five common regulative ones. They furnish its methodological orientation, as a research strategy that pivots on cautious innovation (`Conservatism'), a focus on robust epistemic targets, certain typicality assumptions (`Generalised Copernicanism', in the terms of \textcite{SmeenkWeatherallForthcomingCosmologicalTheory}), a premium on multiple simultaneous motivations (`Common Origin Inferences', \textcite{Janssen2002COIStories}) and a `regulative patrimony' that its gravitational forebears bequeath to today's general-relativistic astrophysics and cosmology. 

This regulative core of the DM paradigm can plausibly be viewed as methodological adaptations to the particular problem situation: the paradigm's regulative principles function as responses to the severe epistemic challenges vexing DM research. To some extent, DM research shares these challenges with cosmology and astrophysics more generally (see, e.g.,\ \textcite{ellis_2006_issues_philosophy_cosmology, DuerrDellsen2025ScientificProgressCosmology, DeBaerdemaeker2025PhilosophyCosmologyAstrophysics}). But also more specific ones can be discerned:

\begin{itemize}
    \item The \textit{elusive nature of DM}---especially its predominantly gravitational effects---limits empirical access, and in particular controlled experimentation \parencite{DeBaerdemaeker2025}. In the same vein, the systems that DM research studies are both spatially and temporally remote, rendering firm epistemic control thorny.
    \item As a result of the foregoing, DM research accrues \textit{sizeable uncertainty} from two sources. First, it ventures into novel and not always fully understood regimes, empirically remote from the contexts in which those laws were initially tested; big leaps in extrapolation are often needed. 
    Secondly, its subject-matter and its astrophysical and cosmological embeddings are complex. They often involve a large nexus of assumptions, models, etc. The evidential chains leading to DM inferences (including highly non-trivial data reduction procedures, alongside astrophysical and cosmological modelling) in particular are long: reasoning about DM (e.g., the inferences from CMB anisotropies to cosmological density parameters) is exuberantly theory-laden.
    \item In DM research \textit{underdetermination} is rampant (see, e.g., \textcite{FerreiraWolfRead2025_SpectreUnderdetermination, Wolf2026}). First, many DM inferences are affected by parameter degeneracies: distinct combinations of model parameters and background assumptions can often reproduce the same observational data. For example, galactic rotation curves may be fitted by varying both the DM distribution and assumptions about stellar mass-to-light ratios, while cosmological observations frequently admit trade-offs among DM density, neutrino masses, and other cosmological parameters. 
    Secondly, underdetermination also arises at a bigger conceptual-theoretical scale---theory underdetermination in the narrower sense, more commonly discussed in philosophy: proposals proliferate that can account for the phenomena in substantively different theoretical ways.   

\end{itemize}

Vis-à-vis this predicament, a cautious, piecemeal strategy for scientific innovation---as opposed to a radically revisionary and, as it were, iconoclastic one---seems prudent. The following five regulative principles put flesh on the bones of such a strategy. As we'll show in \textbf{§\ref{sec:DMcrisis}}, they form natural intra-paradigmatic methodological standards for assessing DM proposals: 
  
\paragraph{Conservatism.} The DM paradigm exhorts researchers to account for DM phenomena in a way that disrupts orthodox physics as little as possible. It counsels what Quine dubbed the ``maxim of minimal mutilation”, but specifically with respect to the principles rehearsed in \textbf{§\ref{ssec:constitutive-principles}}: in devising proposals for DM candidates, keep as many and as much of those principles as possible (cf. \textbf{§\ref{sec:intermezzo}}).

This maxim of minimally mutilating principles receives its normative-epistemic justification as a methodological rule from pragmatic considerations (as Quine himself pointed out, see \textcite[esp.\ ch.\ VI]{QuineUllian1978WebBelief}). First, by drawing on principles (which are, ideally, well-justified in their own right), one mitigates risks incurred by more daring speculations. 
Secondly, such principled inquiry also lowers cognitive costs. As elaborated below, the principles typically contain heuristic resources and fertile links to other domains. They offer valuable guidance for theory or model construction which would otherwise be difficult to obtain---especially if one were to attempt to develop such novel ideas completely from scratch.

To foreshadow some results of our more detailed analysis in \textbf{§\ref{sec:DMcrisis}}:
\begin{itemize}
\item[--] \textit{Sterile neutrinos} require only a minimal extension of the SM’s fermionic sector: they introduce new neutrino types. Apart from this (modest) ontological addition, the SM’s constitutive principles are left intact.
\item[--] Whereas sterile neutrinos extend the SM by enlarging its fermionic particle content, \textit{axions} require a modest extension of the SM’s field content, in tandem with a new global chiral symmetry and associated fields. Apart from this, they leave the SM’s constitutive principles intact, while dynamically modifying aspects of the QCD vacuum structure.
\item[--] Whereas axions extend the SM chiefly through additional fields and a global symmetry, \textit{supersymmetric WIMPs} require deeper structural changes: they extend the Poincaré spacetime symmetry, unify bosons and fermions in supermultiplets, and reconfigure the SM’s quantum-field-theoretic structure, with a host of novel fields and particles.
\item[--] \textit{Primordial black holes} preserve essentially all of the SM’s matter-theoretic and gravitational principles. They shift the burden of new physics almost entirely to (speculative) processes in early-Universe cosmology, and suitable quantum-gravitational effects.

\end{itemize}

Our next regulative principle is a close cognate of Conservatism. It's composed of a cluster of methodological rules  that \textcite{SmeenkWeatherallForthcomingCosmologicalTheory} have  monikered `Generalised Copernicanism' (see also \textcite{Weatherall2021PhilosophyDarkMatter, Weatherall2025SpacetimeModelsCosmology}).

\paragraph{Generalised Copernicanism.} This cluster of ``anti-parochial'' injunctions enjoins us not to epistemically privilege our local spatiotemporal circumstances. It's supposed to ``[make] precise the idea that our place in the world is not special", ``that our position in the world is both \emph{representative}, and in one sense or another, similar to other places" (op.cit.,\ p.\ 10, emphasis in original). 

Most germane to our purposes is `Aristotle's Principle' (\textcite{Weatherall2025SpacetimeModelsCosmology}; see also \textcite{SmeenkBenetreauDupin2017CosmosLocalLaws} for an elaboration). It licences extrapolation of locally established physics, i.e., physics well-tested in our solar system or terrestrial laboratory experiments: ``(l)ocally discoverable laws of physics apply elsewhere in the universe, i.e.,\ in different locations and epochs and, to whatever extent is compatible with known physics, on other length, energy, and time scales" \parencite[p.\ 9]{Weatherall2025SpacetimeModelsCosmology}. Such a regulative postulate checks the unbridled freedom and uncertainty that would otherwise ensue; without it, the bridges to our local knowledge would be severed. The underlying hunch goes, we'd better not touch that Pandora's box of potentially unbounded speculation.\footnote{We can complement this argument from cautious conservatism by one from what Wheeler has called ``\textit{daring} conservatism" (see \textcite{BrillBlum2018DaringConservatism, Furlan2022DaringConservatism}). By teasing out the consequences of our established theories even to the most radical conclusions, we access handy and rich opportunities to \textit{learn}---in particular how far we get with our current knowledge. In this spirit, for instance, \textcite{Efstathiou2023StandardModelCosmology} recommends that in order to advance cosmology and probe the limits of our present understanding, we endeavour ``test $\Lambda$CDM---our `standard model' of cosmology---to destruction."}    

Generalised Copernicanism undergirds---mostly as a tacit\footnote{NB: Already at the \textit{phenomenological} level DM research---whenever we describe and interpret observational data with respect to ramifications for DM---is predicated on Generalised Copernicanism (as is cosmology more generally, see \textcite{ellis_2006_issues_philosophy_cosmology, SmeenkBenetreauDupin2017CosmosLocalLaws, Weatherall2025SpacetimeModelsCosmology}).} premise---the bulk of mainstream DM research. For instance, in constructing models of WIMP freeze-out, mechanisms for axion creation, or the rates of neutrino production, cosmologists extrapolate standard Boltzmann transport equations, quantum statistics, and thermally averaged cross sections from collider physics to epochs with temperatures and densities vastly exceeding terrestrial conditions (see, e.g.,\  \textcite[§5]{BalaczEtAl2024}). By the same token, following their formation, PBHs as DM candidates must be embedded into a cosmic habitat. Treating these processes and interactions with the rest of the universe---prerequisites, after all, for deriving observational consequences---presupposes laws, extrapolated beyond our parochial surroundings.\footnote{Apart from motivating it via principles in its own right, we can also construe the adoption of GR for modelling gravitational effects (\textbf{\S\ref{ssec:gravity-principles}}) as an instance of Copernicanism. Here, it bears reiterating that also on cosmic scales, GR is increasingly well-tested (see, e.g.,\ \textcite{Ishak2019TestingGR}). Conversely, approaches to the DM problem that modify gravity---i.e.\ (MOD-GRAV)---flout it.}

\paragraph{Focus on \textit{robust targets}.} Another countermeasure to the epistemic challenges of DM research is \textit{not} to treat all anomalies on an equal footing, but to home in on phenomena that are reliably and robustly attested to (most impressively whenever many different types of observations consistently point to them, see \textcite{Wimsatt2012RobustnessReliability}). The DM paradigm urges that such targets be prioritised: modelling and explaining them serve as benchmarks on the paradigm's agenda. Less well-controlled, more ambiguous targets should be subordinated---as less urgent or decisive.  

A paradigmatic example for such a robust target is the acoustic peak structure of the CMB power spectrum. Calculations are made in a regime under good control (in particular, the domain where linear perturbation theory is applicable), and the observations, in agreement with those results, are repeatedly confirmed across increasingly powerful missions (COBE, WMAP, Planck). Much of the evidence rehearsed in \textbf{\S\ref{sec:DMparadigm}} fits the mould of this robustness principle. 

Contrariwise, in line with this principle, adherents of the DM paradigm tend not to be overly disconcerted by certain phenomena on galactic scales---where mainstream DM research seems to stall at an impasse:
``(i)t is perhaps not a surprise in this sense that most of the claimed problems of standard cosmology, such as the cusp-core, too-big-to-fail and missing-satellites problems, arise in the deeply nonlinear regime. Model predictions are in this case based on numerical simulations that encode complex processes [...], which is by construction of a potential source of systematic errors" \parencite[p.\ 52]{BertoneTait2018}. In the same vein, \textcite[p.\ 40]{BullockBoylanKolchin2017SmallScale} wrap up their review of the challenges that those galactic (``small-scale") phenomena pose to the DM paradigm: ``(s)mall-scale structure sits at the nexus of astrophysics, particle physics, and cosmology.[..] [T]he level
of agreement between theory and observations remains remarkably hard to assess, in large part because of hard-to-model effects of baryonic physics on first-principles predictions."

Rather than a dogmatic dodge for immunisation (as \textcite{Merritt2020PhilosophicalMOND} castigates), the focus on epistemically secure targets (together with tolerance towards more uncertainty-riddled oddities) reflects the desire to make \textit{genuine} headway, and wariness about wasting time and effort with solutions to what might be artefacts of specious problems. For explorative research on the cusp of the unknown and engulfed by an ocean of bewildering anomalies, this---again thoroughly pragmatic---rationale strikes us as eminently reasonable.

\paragraph{Common Origin Inferences.} A DM approach ought to be well-motivated in that it kills (at least) two birds with one stone (see, e.g.,\ \textcite[p.\ 190]{Feng2010NonWIMP}): it should resolve \textit{other}---at first glance DM-unrelated---problems at once. Linking a DM approach to other problems in this manner is an instance of what \textcite{Janssen2002COIStories} dubbed ``Common Origin Inferences" (COIs) (see also \textcite{DuerrWolfPhilosophies} on COI-reasoning for theory/model construction). COIs denote a historically pervasive inference pattern and a ``powerful engine for theory change" (op.cit.,\ p.\ 470) that ``traces some striking coincidence back to a common origin" (op.cit.,\ p.\ 458). 

What makes COI-reasoning especially suitable for explorative purposes is that it fosters ``fruitful trading zones" \parencite{MartensKing2023_DoingMoreWithLess}. By joining different problems and their respective contexts, COIs permit researchers to tap different cognitive resources. Modelling and theory construction techniques, problem-solving and heuristics can thus be transferred, facilitating the exploration of the DM candidate (cf.\ \textcite{Nyrup2020WaterDrops}).     

The four principal DM candidates all exemplify COI reasoning (see \textbf{§\ref{sec:DMcrisis}} for details):  

\begin{itemize}
    \item[--] \textit{Supersymmetric WIMPs} almost automatically have the DM profile, the right mass density in particular. Supersymmetry, underlying the proposal, is in turn independently motivated as a solution to multiple problems---an instance of COIs in its own right.   
    \item[--] \textit{Axions} were originally proposed as a solution to an apparent fine-tuning problem in quantum chromodynamics. It was subsequently realised that axions would cosmologically evolve like cold DM and could naturally acquire a viable relic abundance.

    \item[--] \textit{Sterile neutrinos} were originally devised to address the origin of neutrino masses (and, in some variants, the baryon asymmetry). By construction, they are stable enough and interact only feebly in any non-gravitational way, as solicited for a DM candidate.   
    \item[--] Interest in \textit{primordial black holes} as DM candidates piqued in the wake of the LIGO/Virgo detection of gravitational waves from merging black holes. Those holes turned out to be somewhat puzzlingly massive. Also the number of merger events subsequently measured exceeded expectations, based on known stellar populations \parencite[§1]{Green2024PBHBriefOverview}. 
\end{itemize}

 \paragraph{Regulative patrimony.} 

The DM paradigm also inherits several regulative principles. In talks reporting forthcoming work \parencite{SmeenkWeatherallForthcomingCosmologicalTheory}, \textcite{Weatherall2021PhilosophyDarkMatter} rightly underlines the methodological continuity between contemporary DM research and the Newtonian and general-relativistic gravitational research programme.

Two of the three regulative principles which Weatherall lists are of primary relevance for our present purposes (and implicitly guide DM mainstream research):
\begin{itemize}
    \item ``no forces without sources”:  gravitational and kinematic phenomena should be attributed to material sources---which further research will hopefully verify; 
    \item the ``action-reaction principle”: such material sources which produce a causal effect suffer a backreaction (as routinely assumed in, e.g., dynamical friction and the response of DM to baryonic matter, especially for DM halos).
\end{itemize}

A pragmatic argument (cf.\ \textcite{Rescher2020MethodologicalPragmatism}) bolsters the force of these inherited regulative principles---as provisional and expedient working hypotheses, as long as they are epistemically fruitful: they proved their mettle in gravitational research, Newtonian\footnote{Over the exemplary applications in celestial mechanics (e.g.,\ \textcite{Smith2014}), one ought not to forget the scarcely less impressive ones in geodesy \parencite{Ohnesorge2024EarthFigure}.} and general-relativistic alike, as Weatherall rightly remarks. Crucially, dispelling a contrary ``myth”, he underscores that \textit{in the context of DM research itself} they have earned their spurs as ``powerful inferential strategies”. In particular, they led to successful predictions (e.g.,\ via gravitational lensing in galaxy clusters, or the first acoustic peak in the CMB spectrum and the ratio between its first and second peak) and coherent explanations (as for the Bullet cluster).

\section{A principled analysis of Dark Matter candidates}\label{sec:DMcrisis}


It's time now to scrutinise the primary DM candidates within the DM paradigm---supersymmetric WIMPs (\textbf{\S\ref{sec:WIMPs}}), axions (\textbf{\S\ref{sec:axioms}}), sterile neutrinos (\textbf{\S\ref{sec:sterileneutrinos}}), and primordial black holes (\textbf{\S\ref{sec:PBH}})---and where they stand with respect to principled inquiry.

\subsection{Supersymmetric WIMPs}\label{sec:WIMPs}

Supersymmetry (`SUSY'\footnote{For physics introductions, see e.g.\ \textcite{WessBagger1992, Aitchison2007, BailinLove1994, Terning2006}; for philosophical discussions see in particular \textcite{Menon2021-MENTUS-2} as well as the topical collection of \textcite{MenonTC} and articles therein.})
is a hypothesised symmetry relating bosons and fermions. The SM segregates bosons (i.e., particles with integer spin, including force carriers such as photons or gluons) and fermions (i.e., particles with half-integer spin, including matter particles such as electrons and quarks) into disparate classes. With SUSY, bosons and fermions are united within a common structure.\footnote{In terms of representation theory, bosons and fermions in the SM are separate irreducible representations of the Poincaré group, whereas in SUSY they become irreducible representations---`supermultiplets’---of the super-Poincaré algebra, containing \textit{both} bosonic and fermionic states.} SUSY implies the existence of a panoply of `superpartners', one for each of the SM particles. While none of those superpartners---or other signs of SUSY---has hitherto been detected, it’s tempting to view them as premier potential DM candidates.

As a \textit{mathematical} possibility, SUSY was discovered in the early 1970s, in particular with the work of \textcite{Golfand:1971iw} in the Soviet Union and, independently, by \textcite{Wess:1974tw} in the West.\footnote{For more detailed histories of SUSY for philosophical audiences, see \textcite{Castellani2026Supersymmetry}, \textcite[§2]{HetzroniRead2026}, and \textcite[§2]{MenonTC}.} Physicists had been seeking to unify spacetime symmetries with internal symmetries (i.e.\ symmetries acting on particles’ internal degrees of freedom)---quantum numbers in particular---rather than on spacetime coordinates, such as the $U(1)$ gauge symmetry of electromagnetism. Early results---most notably a theorem due to \textcite{ColemanMandula}---seemed to forbid non-trivial mixing of spacetime and internal symmetries, except in very restricted ways.
The breakthrough which led to SUSY was the realisation that if one allows not only commuting generators (i.e., the operators that generate the symmetries) but also anti-commuting generators, then the conditions of the Coleman--Mandula theorem can be bypassed.

This ostensibly modest \textit{mathematical} generalisation revealed an unexpected \textit{ontological} ramification: whereas ordinary symmetry generators---those of the SM---transform particles only within a given class (i.e., transforming one bosonic field into another, or one fermionic field into another), the new generators map bosonic field states into fermionic ones, and vice versa. 
To realise this enlarged symmetry, the SM itself must be reconfigured so that its particle content, interactions, and dynamical laws are all invariant under these new transformations. In particular, every known particle must be incorporated into a larger supermultiplet together with a superpartner (whose mass---in the realistic case of broken SUSY (see below)---depends on the parameters governing SUSY breaking, which in turn are determined by empirical constraints). In the same vein, the interaction terms of the theory must be modified accordingly. The familiar SM thus emerges as part of a more comprehensive supersymmetric QFT.

Through the late 1970s and 1980s, SUSY swiftly ascended to a major research programme for high-energy physics (see \textcite{HaberKane1985} for details). Its appeal derived from SUSY’s promise to address several otherwise independent theoretical issues bedevilling the SM (see also \textcite{Fischer2024PromiseOfSupersymmetry}): 
\begin{itemize}
\item The most influential motivation for low-energy SUSY was the hierarchy problem, associated with the Higgs boson and commonly understood as a problem of naturalness or fine-tuning (see, e.g.,\ \textcite{Fischer2023Naturalness}). Within the SM understood as an effective field theory, the Higgs mass receives quantum corrections from its interactions with other fields. If the SM remains valid up to a high cut-off---often taken to be the Planck scale, where the onset of gravitational effects is expected---the Higgs mass parameter receives corrections many orders of magnitude larger than the observed Higgs mass. Explaining the observed mass would require a suspiciously fine-tuned cancellation between the bare Higgs mass and these quantum (`radiative') corrections.
SUSY offers a mechanism to avoid this fine-tuning. Bosonic and fermionic quantum corrections enter with opposite signs. The dominant contributions to the Higgs mass hence cancel between each particle and its superpartner. Consequently, the Higgs mass remains stable against radiative corrections, \textit{without} requiring delicate cancellations between independent parameters.
The cancellation is exact only for \textit{full} SUSY, however. As no superpartners have been observed yet, SUSY must be broken: although SUSY remains part of the underlying theory, the vacuum state and resulting particle spectrum no longer show it exactly. The effectiveness of SUSY-induced cancellations of quantum corrections depends on the mass splitting between SM particles and their superpartners. If the latter are too heavy, the residual corrections to the Higgs mass again become large, reintroducing the fine-tuning SUSY was meant to alleviate. Naturalness therefore suggested that at least some superpartners should appear near the electroweak scale. This provided a central incentive for collider searches, such as at the LHC---and rendered SUSY especially tantalising for DM (see below).

\item A second motivation stems from the \textit{possibility of Grand Unification}. The SM describes the strong, weak, and electromagnetic interactions through three separate gauge symmetries. Each is associated with a gauge coupling whose value changes (`run') with energy. When these couplings are extrapolated to very high energies, they approach one another but fail to meet \textit{exactly}. This impedes a unification of the three forces into a single underlying interaction---a powerful theoretical desideratum driving high-energy physics (see, e.g.,\ \textcite{Salimkhani2018QuantumGravity, Salimkhani2021ExplainingUnification}).
Supersymmetric extensions of the SM modify the running of the gauge couplings. Thereby the couplings can converge remarkably closely at about $10^{16}$GeV. In the eyes of SUSY advocates, this near-intersection boosts confidence in SUSY as a plausible low-energy remnant of a more fundamental grand unified theory. 
\item SUSY turned out to be an indispensable ingredient of \textit{superstring theory}. It removes the tachyonic instability of the bosonic string, yields a consistent spectrum containing both bosons and fermions, and allows gravity to emerge naturally through the appearance of a massless spin-2 graviton. While superstring theory remains experimentally unconfirmed, its dependence on SUSY reinforced the view that SUSY reflects a deeper principle of fundamental physics rather than an ad hoc add-on of the SM.
\item While the SM---after the demise of `hot' DM in the form of (ordinary) neutrinos in the mid/late 1980s---contains no particles with the requisite properties of DM candidates, SUSY enlarges the particle spectrum in ways that include particles matching the phenomenological profile inferred for DM: supersymmetric WIMPs. 
\end{itemize}
It’s on this last point that we’ll now zoom in.
SUSY WIMPs have enjoyed special popularity as DM candidates because they arise almost automatically in supersymmetric extensions of the SM with the right profile (see \textcite{Roszkowski2018,Catena2014,Baer2016,Jungman1996}, whom we follow): 
\begin{enumerate}
\item[(1)] \textit{Stability.} To escape conflicts with experiments, many SUSY models impose a discrete symmetry known as `R-parity’. Every SM particle is assigned an R-parity of $+1$, every superpartner an R-parity of $-1$. Parity conservation forbids certain interactions---in particular, those that would threaten SUSY’s empirical adequacy (by violating baryon and lepton number and rendering processes such as proton decay far too rapid). If R-parity is conserved, superpartners can only decay into other superpartners (plus ordinary particles); the lightest supersymmetric particle (LSP) cannot decay into SM particles \textit{alone}. Since no lighter supersymmetric state exists into which it could decay, the LSP is absolutely stable. Unlike ordinary unstable particles, it can therefore survive from the early Universe until the present day.
\item[(2)] \textit{Weakly-interacting and electrically neutral.} These features don’t follow from SUSY alone, but from the most studied phenomenological models, most notably the Minimal Supersymmetric Standard Model and closely related models. In them, the LSP is a neutral fermion known as the neutralino. It’s a superposition of the superpartners of the neutral gauge bosons (the bino and wino) and the neutral Higgs bosons (the higgsinos). Because it carries no electric charge, the neutralino doesn’t interact electromagnetically. Its interactions are instead mediated primarily through the weak interaction (and, depending on its composition, Higgs exchange). 

\item[(3)] \textit{Density/abundance.} In the early Universe, a particle with weak-scale mass and interaction strength comparable to the weak force naturally enters thermal equilibrium and subsequently undergoes `freeze out' when its annihilation rate drops below the Hubble expansion rate. Intuitively, the Universe expands faster than particles bump into each other to annihilate. Their abundance `freezes' to a constant value per comoving volume. The relic abundance produced in this process is remarkably close to the observed DM density. This coincidence---that a weak-scale particle can give the right order of magnitude for the DM abundance---is called the `WIMP miracle'.
\end{enumerate}
Next, let’s supplement this more standard motivation of SUSY WIMPs with an analysis through the prism of principles (\textbf{§\ref{sec:principlesofthedmparadigm}}):
\paragraph{Constitutive principles} The introduction of SUSY WIMPs doesn’t merely amount to extra particle species. Rather, supersymmetric extensions modify several of the SM’s constitutive principles:\footnote{For the purposes of dark matter phenomenology it is sufficient to consider supersymmetric extensions of the Standard Model. If one wishes to implement SUSY as a fundamental symmetry also of spacetime---a possibility that appears quite cogent given SUSY’s mixing of spacetime and internal symmetries---one is led to the development of a supergravity theory. In that case, many constitutive principles of gravity are modified. Since the move to supergravity isn’t usually countenanced in the context of discussions of SUSY WIMPs, we set aside this fairly drastic departure from established physics.}
\begin{itemize}
\item \textit{From Poincaré to super-Poincaré invariance.} SUSY’s most fundamental modification concerns the SM’s reliance on Poincaré invariance as a spacetime symmetry. SUSY extends the Poincaré algebra to the larger super-Poincaré algebra. As sketched above, this necessitates deep structural revisions of the SM. 
\item For conformity with SUSY, one must \textit{reorganise the SM’s quantum field-theoretic structure}. Every SM particle, in particular, must be embedded into a supermultiplet, containing a superpartner. Care must be taken in choosing suitable interaction terms. For this, one typically postulates R-parity (see above) as a further adequacy constraint (in the sense of \textbf{\S\ref{ssec:principledinquiry}}). 
As mentioned, the structural modifications also have ontological consequences. First, SUSY enlarges the SM’s physically admissible particle spectrum of the observed quarks, leptons, gauge bosons, and Higgs degrees of freedom: the minimal supersymmetric extension roughly doubles the number of fundamental degrees of freedom. Secondly, bosons and fermions cease to be fundamentally separate categories; instead, they are fused within supermultiplets. 
\item By the same token, SUSY preserves the SM's \textit{gauge and chiral structure} while embedding both within supersymmetric multiplets.
\item Finally, SUSY enhances \textit{renormalisability}. SUSY doesn’t convert an otherwise non-renormalisable theory into a renormalisable one. But it strengthens the ultraviolet (high-energy) behaviour of renormalisable QFTs by severely constraining the form of quantum corrections.\footnote{These restrictions become even more stringent in theories with extended supersymmetry. For example, $\mathcal{N}=4$ supersymmetric Yang--Mills theory is perturbatively finite to all orders.} As indicated above, as part of SUSY’s promise to resolve the Higgs naturalness problem, this secures the stability of the supersymmetrised QFTs under quantum corrections without requiring delicate fine-tuning.
\end{itemize}
\paragraph{Regulative principles}.
\begin{itemize}
\item \textit{Conservativism.} We saw above how the extension of Poincaré algebra to the SUSY algebra as a fundamental symmetry entailed a thorough and highly non-trivial re-working of the SM. Moreover, and in particular, SUSY introduces an entire suite of new particles and interactions. While (to stay in the image of Quine’s metaphor, \textbf{\S\ref{sec:intermezzo}} and \textbf{\S\ref{ssec:regulative-principles}}) not mutilating the SM’s principles---in not tearing them apart wholesale---SUSY quite mercilesssly stretches them; strenuous efforts are needed to prevent fissures (e.g., adjustment of parameters beyond what naturalness would licence or the introduction of R-parity). 
\item \textit{Robustness.} Experimental attempts to detect SUSY have significantly shrunk the viable parameter space. Prospects of empirical detection of SUSY particles in the near future look dim: ``(f)or more than fifteen years of LHC operation, the CMS and ATLAS collaborations have achieved remarkable sensitivity to a wide range of supersymmetric signatures. Despite this unprecedented reach, no conclusive evidence for supersymmetry has emerged. If supersymmetry is nature’s solution to outstanding questions in particle physics, it is necessarily challenging to find’’ \parencite[p.\ 1]{JeantyLee2026RareSUSY}. Given this status quo, the focus has shifted away from robust targets (cf.\ \textcite{King2025}).\footnote{\label{fn:MSSM}The Minimally Supersymmetric Standard Model has 124 free parameters---105 more than the SM. This large number of degrees of freedom evidently butts against robustness.}
\item \textit{Generalised Copernicanism} was initially, in light of the WIMP miracle, one of the strongest arguments favouring SUSY WIMPs: researchers extrapolated generic weak-scale particle physics measured in terrestrial experiments to the thermal history of the early Universe. Current observational constraints, however, spoil the Copernicanism: satisfying collider and direct-detection limits pushes SUSY WIMPs into ever more fine-tuned pockets of parameter space (see also ibid.;\ \textcite[§5]{Baer:2025zqt}). Vis-à-vis Generalised Copernicanism SUSY WIMPs are best seen as (at best) a mixed bag.
\item \textit{COI inferences.} As we saw, SUSY’s enormous appeal lay in its confluence of a number of other claimed merits \textit{at once}. As such, SUSY WIMPs constitute a case of a COI inference \textit{par excellence}: SUSY WIMPs remain strongly motivated on independent (theoretical) grounds. 
That said, the requisite fine-tuning for SUSY’s viability has diminished its claims to naturalness (ibid.). But most of the other motivations still stand.
\item \textit{Regulative patrimony.} Proponents of SUSY may be inclined to point to some methodological continuity in the transition from non-SUSY to SUSY field theories: extending symmetry groups seems to mimic a standard move for successful theory development in contemporary physics (cf.\ \textcite{Castellani2026Supersymmetry, Friedan1986Conformal}). 
However, SUSY lacks a convincing rationale for such an appeal to regulative parsimony. As \textcite{Hetzroni2025Projectability} urges, the crucial question is whether the transition can be understood as an extrapolation from local empirical evidence: does SUSY project what is responsible for previous theories’ empirical success? In contrast to, say, the move from the Lorentz symmetries of SR to the diffeomorphism group of GR \parencite{HetzroniForthcoming-HETHTT-2}, the answer here is `no’: SUSY is devoid of direct \textit{empirical} warrant. While displaying the formal pattern of enlarging the symmetry group, SUSY does so in a highly \textit{speculative} way, not grounded in established successes analogous to those underwriting earlier symmetry extensions (see \textcite{HetzroniRead2026} for further discussion).
\end{itemize}
Overall, principled inquiry yields a less favourable appraisal of SUSY WIMPs than their historical prominence in DM research would suggest. They require a profound restructuring of the SM’s constitutive principles; compatibility with present constraints depends on delicate theoretical engineering, and ever more artificial fine-tuning. With regard to the DM paradigm’s regulative principles, SUSY WIMPs fare even more poorly. Only COI reasoning continues to offer substantial---albeit somewhat weakened---motivation. SUSY WIMPs receive a low score on conservativism, increasingly fail robustness, and sit uneasily with Generalised Copernicanism. As far as principled inquiry is concerned, their historically privileged prioritisation ought to be renegotiated.\footnote{To better appreciate the tension between our evaluation and the popularity that SUSY WIMPs have historically enjoyed we conjecture that the latter is owed to the firm expectation that the days of the SM were counted. The SM was widely believed to be superseded \textit{soon}, with hints of Beyond SM physics just around the corner. SUSY was deemed an exceptionally auspicious element of such Beyond SM physics.

\textit{Today}, after decades of increasingly stringent null results, optimism about Beyond SM physics has plummeted (see, e.g.,\ \textcite{Hossenfelder2018, Hossenfelder2021Screams}); scepticism about SUSY in particular is understandably growing. In this changed epistemic situation it strikes us as reasonable to ascribe to the SM, warts and all, methodological default status---a status it quite legitimately lacked twenty years ago---and proceed cautiously from \textit{its} principles. That is, given the SM's enduring empirical success and the absence of evidence for Beyond SM physics (with concomitantly dwindling trust in naturalness as a reliable heuristic principle, ibid.), epistemic caution counsels conservatism with respect to \textit{the SM’s} principles.}

\subsection{Axions}\label{sec:axioms}

Axions are hypothetical particles originally proposed as a solution to the \textit{strong CP problem} of quantum chromodynamics (QCD) (see, e.g.\, \textcite{KimCarosi2010,Marsh2016} for details; and \textcite{Dougherty} for a philosophical analysis of the CP problem specifically). To understand their origin, recall that the SM is built from QFTs whose interactions are strongly constrained by symmetry principles. QCD---the theory describing the strong nuclear force---is a gauge theory based on the symmetry group $SU(3)$ and describes the interactions of quarks and gluons, with the latter as the gauge bosons mediating the force between quarks.
The symmetry (and other) principles underlying QCD permit---that is, they do not \textit{per se} forbid---a term of the form\footnote{Cf.\ our discussion of the Principle of Totalitarianism below and in \textbf{\S\ref{sec:sterileneutrinos}}.}
\begin{equation}\label{eq:theta}
\mathcal{L}_{\theta}=\theta\frac{g_s^2}{32\pi^2}G_{\mu\nu}\widetilde{G}^{\mu\nu}
\end{equation}
in the QCD Lagrangian, with $g_s$ the strong coupling constant (setting the strength of the quark--gluon interaction). Here, $G_{\mu\nu}$ denotes the gluon field strength tensor, describing the intensity and structure of the gluon field, closely analogous to the electromagnetic field strength tensor. As in electromagnetism, one may define its dual tensor $\widetilde G_{\mu\nu}$ by swapping the electric- and magnetic-like components of the field.

The (dimensionless) CP parameter $\theta$ controls the strength of this additional gluon self-interaction. It determines the extent to which the strong interaction violates the combined symmetry of charge conjugation $C$ (replacing particles by their antiparticles) and parity $P$ (reflecting spatial coordinates), known collectively as CP symmetry. In particular, a non-vanishing $\theta$ term induces a permanent electric dipole moment for the neutron---a separation between the centres of its positive and negative charge distributions (carried by the up quark and the two down quarks, respectively). Like the electric dipole moment of a water molecule, such a dipole would reveal itself through its interaction with an external electric field, most notably by producing a tiny electric-field-dependent shift in the neutron's spin precession.

So far, experiments searching for the neutron electric dipole moment have found no such effect. They place extremely stringent upper bounds on its possible size: the physical value of $\theta$ must be extraordinarily small.\footnote{Strictly speaking, the physical parameter is not simply the bare parameter $\theta$, but a combination involving phases in the quark mass matrix.} Should we accept this at first glance puzzlingly special numerical value as a brute fact, or does it scream for a deeper explanation? The gist of the `strong CP problem' concerns precisely this apparent fine-tuning.

The Peccei--Quinn (PQ) mechanism proffers such an explanation by introducing a new dynamical field whose coupling to QCD promotes the effective strong CP parameter from a fixed constant to a dynamical quantity.\footnote{The original papers by \textcite{PecceiQuinn1977PRL,PecceiQuinn1977PRD} established that this enlarged symmetry structure dynamically drives the vacuum towards a CP-conserving state, thereby explaining the observed smallness of the effective strong CP parameter rather than merely postulating it. \textcite{Weinberg1978,Wilczek1978} subsequently recognised that the new dynamical degree of freedom implied by the mechanism should manifest itself as a physical light pseudoscalar particle---the axion---whose interactions could in principle be sought experimentally.}

More precisely, the PQ mechanism enlarges the SM by dint of entirely standard methods for constructing QFTs. First, one introduces an additional global continuous symmetry, denoted $U(1)_\text{PQ}$, together with a new complex scalar field, $\Phi(x)$, on which the PQ symmetry acts. 
As is customary in QFT---and chiming with what \textcite{GellMann} dubbed the (regulative) `Principle of Totalitarianism’ (sometimes, and more aptly, also called the `Principle of \textit{Plenitude}’, see \textcite{Kragh2019}, \textcite{Kragh2019Plenitude} for historical reviews\footnote{The Principle---the licence to include every term consistent with the theory's symmetries unless there is a symmetry that forbids---is a standard heuristic in theory and model building within Effective Field Theory.})---the scalar is assumed to be endowed with the most general renormalisable self-interaction compatible with the assumed symmetries: the Mexican-hat potential, familiar from the Higgs formalism. Its minimum doesn’t occur at $\Phi=0$, but rather on a circle of degenerate minima satisfying
$|\Phi|=f_a/\sqrt2$.

The field configuration therefore settles into, or chooses at random, one particular vacuum from this continuous family of equally energetic possibilities---a phenomenon known as spontaneous symmetry breaking: although the underlying equations retain the full $U(1)_\text{PQ}$ symmetry, the vacuum state itself does not.\footnote{The standard analogy is a perfectly symmetric pencil balanced on its tip. The laws governing the system are rotationally symmetric. But once the pencil falls, it picks a particular direction---thereby breaking that symmetry. For more philosophy of physics work on spontaneous symmetry breaking, see e.g.\ \textcite{sep-symmetry-breaking, earman2003rough} and references therein.}
More generally, Goldstone's theorem guarantees that, in relativistic QFTs, whenever a continuous global symmetry is spontaneously broken, one of the field’s degrees of freedom appears as a new massless scalar excitation. This so-called Goldstone boson corresponds to fluctuations that move the system from one degenerate vacuum to a neighbouring one (`horizontally' in the circle of vacua, as it were) at no energy cost. One may parametrise the scalar field after symmetry breaking as
\begin{equation}
\Phi(x)=\frac{1}{\sqrt2}\bigl(f_a+\rho(x)\bigr)e^{ia(x)/f_a}.
\end{equation}
Here, $\rho(x)$ describes fluctuations in the radial direction; $a(x)$ describes fluctuations in the phase. The scale of the PQ symmetry breaking is $f_a$. The radial mode typically has a mass of order $f_a$, whereas the phase field remains massless at the classical level. This Goldstone boson is called the \textit{axion}.
At the classical level the theory is invariant under shifts of the Goldstone field,
\begin{equation}
a(x)\rightarrow a(x)+\text{constant},
\end{equation}
reflecting the fact that all points on the circle of degenerate vacua are energetically equivalent.
The new symmetry is, however, not an exact symmetry of the \textit{quantum} theory. Although the classical Lagrangian is invariant under the $U(1)_\text{PQ}$ symmetry, quantum fluctuations of the quark fields spoil that invariance: certain loop effects render the otherwise conserved PQ current non-conserved, a phenomenon known as a quantum anomaly. Intuitively, the symmetry survives at the level of the classical equations of motion but fails once quantum fluctuations of the fields are taken into account.\footnote{Recall that every QFT begins with a classical field theory described by a Lagrangian, whose equations of motion determine the dynamics of classical fields. Quantisation promotes those classical fields to quantum operators and incorporates their fluctuations. Although symmetries of the classical Lagrangian often survive this procedure, some do not. In some cases quantisation obstructs the conservation law associated with a classical symmetry. Such a symmetry is said to possess a quantum anomaly---as in the PQ symmetry.}

This anomalous breaking is an essential ingredient deliberately built into the Peccei--Quinn mechanism. The particle content and symmetry assignments are chosen precisely so that the PQ symmetry possesses the required QCD anomaly. Were the symmetry exact also quantum mechanically, the Goldstone boson would remain exactly massless and completely decoupled from the CP-violating QCD interaction. Instead, the anomaly gives the Goldstone field precisely the required coupling to the gluonic self-interaction term above,
\begin{equation}
f^{-1}_aa(x)G_{\mu\nu}\widetilde G^{\mu\nu},
\end{equation}
which has exactly the same form as the original $\theta$ term. The physically relevant coefficient of the CP-violating interaction becomes
\begin{equation}
\theta_\text{ eff}=\theta+a(x)/f_a,
\end{equation}
rather than the constant parameter $\theta$ alone. The original parameter $\theta$ remains a free parameter of the underlying theory, but the effective coefficient governing CP violation has become dynamical.

The axion isn’t exactly massless, but a so-called \textit{pseudo}-Goldstone boson; although the Peccei--Quinn shift symmetry is broken by the QCD anomaly, it’s the nonperturbative dynamics of QCD that generates the axion potential and mass, lifting the otherwise flat direction.
In other words, the potential is no longer perfectly flat around the circle of degenerate vacua: it develops a small curvature near its minimum. This curvature determines the axion’s mass, giving the axion a tiny but finite mass. Unlike the radial mode, whose mass is set by the scale of Peccei--Quinn symmetry breaking (arising from the curvature of the Mexican-hat potential in the radial direction), the axion mass is generated only through non-perturbative QCD effects.

These occur at low energies---as physical effects that cannot be captured by an expansion in powers of the strong coupling constant (because, at low energies, the strong interaction is simply too strong for perturbation theory to be reliable). They generate an additional potential for the axion field (on top of the above self-interaction potential, responsible for the spontaneous symmetry breaking). It lifts the otherwise perfectly flat circular valley of degenerate minima, selecting one particular point around the circle as the \textit{true} vacuum.

In terms of the axion field, this means that the previously flat direction around the circle is no longer energetically uniform: different values of the axion field now correspond to different vacuum energies. The observable CP violation depends only on the combination $\theta_\text{ eff}=\theta+a(x)/f_a$. The effective potential has its global minimum at $\theta_\text{eff}=0$.

The axion field evolves according to the equations of motion derived from this potential. For a broad range of initial conditions, it `rolls’ towards the minimum of the potential. Eventually, it settles there once its oscillations are damped by the expansion of the Universe (or, more abstractly, because the vacuum state corresponds to the lowest-energy configuration of the theory). In this equilibrium state,
\begin{equation}
\theta+f^{-1}_a\langle a\rangle=0.
\end{equation}
The effective CP-violating interaction is thereby dynamically relaxed to zero. 
The strong CP contribution to the neutron electric dipole moment is thereby effectively suppressed.
In this way, the Peccei--Quinn mechanism explains the observed smallness of the effective strong CP parameter by making the CP-conserving value the \textit{dynamically} preferred vacuum configuration.

In 1983, shortly after the axion’s initial proposal in the late 1970s, three research groups independently realised that axions would make for an in-principle viable DM candidate: the same dynamical evolution that relaxes the effective $\theta$-parameter also implies a cosmological evolution of the axion field that naturally leaves behind a relic population mimicking cold DM.

Unlike WIMPs and sterile neutrinos, axions needn’t be produced as individual particles in collisions in the hot primordial plasma. Instead, their dominant production mechanism is believed to be `vacuum misalignment’---the coherent motion of the axion field itself (i.e., the nearly homogeneous axion field oscillating in sync about the minimum of its potential).

To understand this process, recall that the axion field is created when the PQ symmetry is spontaneously broken. The field then assumes at random a definite value, corresponding to one point on the circle of degenerate vacua. The axion field generally starts out displaced from its eventual equilibrium value (i.e., the CP-conserving minimum in units of the PQ symmetry-breaking scale).

Initially, this displacement has negligible dynamical consequences. In the very early Universe the expansion rate is extremely large, so that the rapid expansion effectively damps the evolution of the field. The equation governing the homogeneous axion field contains a friction term proportional to the Hubble expansion rate,
\begin{equation}
\ddot a+3H\dot a+\frac{\partial V(a)}{\partial a}=0,
\end{equation}
where $H:=\frac{dR(t)/dt}{R(t)} $ denotes the Hubble parameter (with the scale factor $R(t)$). At early times this `Hubble friction' dominates over the restoring force arising from the potential; the field remains almost frozen at its initial value, despite \textit{not} lying at the minimum of the potential.

This situation changes as the Universe expands. Its expansion rate steadily decreases, reducing the Hubble friction; the previously frozen axion field `thaws’. At the same time, as the Universe cools and enters the regime where QCD becomes strongly coupled, non-perturbative QCD effects generate the axion potential discussed above.

The emergence of this potential lifts the previously flat direction of the axion field and provides a restoring force that drives the axion towards its CP-conserving minimum. The axion begins oscillating coherently about the minimum of the potential, in much the same way that a pendulum released from rest oscillates around its equilibrium position.

Remarkably, although we are \textit{not} dealing with particles but coherent fields (equivalently, with a quantum state containing an enormous occupation number of extremely low-momentum axions), their oscillations, on cosmological time scales, behave like ordinary non-relativistic dust particles. Averaged over many oscillations, the energy density stored in the field is diluted by cosmic expansion like $\rho_a\propto R_{\text{scale}}^{-3}$---exactly the same dependence exhibited by cold DM. 

The amount of DM produced depends primarily on the axion mass, the PQ scale $f_a$, and the initial misalignment angle $\theta_i$. For suitable values of these parameters, the energy stored in the initially displaced axion field evolves naturally into a relic abundance comparable to the observed DM density. Note that for the QCD axion, the mass and interaction strengths are not independent parameters: the axion mass $m_a$ is determined by the PQ scale $f_a$ through the QCD anomaly, with larger $f_a$ corresponding to lighter axions and weaker couplings, roughly $m_a \propto f_a^{-1}$ .

The cosmological history is sensitive to \textit{when} the PQ symmetry breaks. If it happens before cosmic inflation, inflation stretches one nearly homogeneous value of the axion field across the observable Universe. In that case, the relic abundance depends on a single initial misalignment angle. If instead the symmetry breaks after inflation, different regions of the Universe settle into different points on the circle of vacua. The subsequent evolution becomes considerably richer, involving axion strings, domain walls, and additional axion production from their decay. Although these scenarios differ quantitatively, both illustrate the same underlying principle: the DM abundance emerges from the cosmological evolution of a field introduced on entirely independent particle physics grounds.\

The foregoing shows that in principle it’s \textit{possible} for axions to comprise DM. But do physically realistic axion theories exist that naturally possess the required combination of masses, couplings, and cosmological histories? For this, we need to inspect concrete axion models that arise from the PQ mechanism and its generalisations. For our purposes, we may confine ourselves to the two main ones: the KSVZ and the DFSZ model, respectively.

The Kim--Shifman--Vainshtein--Zakharov (KSVZ) model is one of the simplest and most widely studied realisations of the PQ mechanism. It specifies a minimal extension of the SM whose particle content possesses the QCD anomaly required for the PQ mechanism.

Besides the SM fields, the KSVZ model posits two new fields: the complex scalar field $\Phi$ (whose phase will be identified with the axion after $\Phi$’s spontaneous PQ symmetry breaking) and a new heavy vector-like quark $Q$. That is, unlike the ordinary quarks of the SM (whose left- and right-handed components transform differently under the electroweak interaction), both of $Q$’s chiral components transform identically. 

Because $Q$ is vector-like, the SM gauge symmetries \textit{alone} would permit---be consistent with (heuristically in line with the Principle of Totalitarianism)---a mass term, quadratic in the quark fields,
\begin{equation}
m_Q\overline QQ.
\end{equation}
Next, we impose PQ symmetry (employing it a constitutive principle) in order to forbid this otherwise-allowed term, so that the heavy quark mass originates dynamically---as an \textit{effective} mass---from PQ symmetry breaking, as sketched above.

The PQ symmetry, in tandem with the other symmetry principles and renormalisability in the SM, allows for another term, however: the Yukawa interaction,
\begin{equation}
\mathcal{L}_\text{Yukawa} = y_Q\Phi \overline Q_LQ_R+\text{Hermitian conjugate},
\end{equation}
where the dimensionless coupling constant $y_Q$ determines the interaction strength between the quark and the PQ scalar. 
This Yukawa term describes the heavy quark’s coupling to the scalar field: a left-handed quark can be converted into a right-handed one by emitting or absorbing a quantum of $\Phi$, and vice versa (the latter process being represented by the Hermitian conjugate). 

The Yukawa interaction is integral to the KSVZ model: once $\Phi$ acquires a non-zero vacuum expectation value through spontaneous PQ symmetry breaking, replacing $\Phi$ by its vacuum expectation value transforms the Yukawa interaction into an ordinary mass term (the direct route to which, via the explicit mass term above, was blocked by the PQ symmetry),
 \begin{equation} 
m_Q = y_Q\langle\Phi\rangle \simeq y_Qf_a. 
\end{equation}
In other words, $Q$ acquires its mass thanks to the spontaneously broken PQ symmetry (quite analogously to the way the spontaneously broken Higgs field generates the masses of the SM fermions).

At the same time, $\Phi$’s \textit{phase}---the axion, as we’ll spell out below---remains coupled to the heavy quark via the Yukawa interaction. Astrophysical observations (e.g., from stellar cooling) and laboratory experiments mandate that the PQ symmetry-breaking scale $f_a$ lie far above experimentally accessible energies. Unless we are willing to swallow a suspiciously fine-tuned value of the coupling constant $y_Q$ (or, contrariwise, if we adopt the (regulative) principle of naturalness, recall \textbf{\S\ref{sec:WIMPs}}), the quark’s mass $m_Q =y_Q f_a$ far exceeds the (comparatively low) energies of the processes of interest. Its effects can therefore, in the parlance of Effective Field Theory, be `integrated out': we may remove the heavy quark as an explicit degree of freedom; its effects are absorbed by effective interactions amongst the remaining fields. The quark’s contribution to the effective axion--gluon interaction term due to the QCD anomaly, however, survives. It’s that interaction term upon which PQ mechanism pivots in the KSVZ model.

What, finally, of the axion? The KSVZ quark’s \textit{raison d’être} is to introduce new matter content needed for the PQ symmetry to acquire the QCD anomaly on which the mechanism relies. Upon quantisation, this anomaly generates a coupling between the axion and the gluon field. Non-perturbative QCD effects can then, via the PQ mechanism, generate the axion potential—giving the axion a small mass, approximately $m_a\propto f_a^{-1}$ (whilst dynamically driving the effective strong CP parameter to zero).
Without the heavy quark, this anomalous coupling would be absent; the axion would remain a Goldstone boson---and as such massless. 

Independent astrophysical observations require the PQ scale $f_a$ to be very large. Were the axion coupled too strongly to ordinary matter, it would provide an efficient channel through which stars could lose energy. This would substantially alter stellar evolution. Likewise, the neutrino signal observed from Supernova~1987A would have been noticeably different. These considerations already imply $f_a\gtrsim10^{8\text{--}9}\,\mathrm{GeV}$---placing the KSVZ quark far beyond the reach of present-day particle accelerators.

Crucially for the purposes of our paper, the region of parameter space consistent with these independent particle-physics and astrophysical constraints overlaps with that for which the vacuum-misalignment process produces approximately the observed DM abundance! For a generic initial misalignment angle of order unity (as naturalness would make us expect), this occurs for $f_a\sim10^{11}\text{--}10^{12}\,\text{GeV}$, corresponding to an ultra-light axion mass of about $ m_a\sim10^{-5}\text{--}10^{-4}\,\text{eV}$ (\textit{significantly} below the currently known neutrino mass scale).

A second realisation of the PQ mechanism is the Dine--Fischler--Srednicki--Zhitnitsky (DFSZ) model. The detailed particle content differs (with slightly different empirical signatures in this regard), but the main building blocks---specifically the central role played by the PQ symmetry and its QCD anomaly---are retained. Like its KSVZ rival, the DFSZ model predicts an ultralight axion. The key difference lies in how the QCD anomaly, required for the PQ mechanism (as responsible for the axion's coupling to the gluon field), is brought about. Rather than introducing a new heavy quark whose PQ transformation properties generate the anomaly, the DFSZ model incorporates the PQ symmetry into the ordinary SM quarks themselves. To make this enlarged symmetry compatible with the interactions responsible for the SM’s fermion masses, the scalar sector (i.e., the collection of scalar fields) must likewise be enlarged.

More precisely, in addition to the SM’s fields (including the ordinary Higgs field), the DFSZ model posits a second Higgs doublet (that is, a second Higgs field with the same electroweak symmetry properties as the ordinary Higgs) and a complex scalar field $\varphi$. The latter is responsible for spontaneous PQ symmetry breaking. The principal role of the second Higgs doublet is to allow the ordinary Yukawa interactions (responsible for generating quark and lepton masses) to remain compatible with the imposed PQ symmetry. Without enlarging the Higgs sector in this way, introducing the PQ symmetry would either forbid some of those mass-generating interactions or require more expansive modifications of the SM.

As in the KSVZ model, the scalar $\varphi$ acquires a non-zero vacuum expectation value,
\begin{equation}
\langle\varphi\rangle\sim f_a,
\end{equation}
spontaneously breaking the PQ symmetry. As before, this produces an axion as the associated pseudo-Goldstone boson. In contrast to the KSVZ model, the required QCD anomaly now arises from \textit{ordinary} quarks themselves, rather than from a newly introduced fermion. This anomaly generates the same effective axion–gluon coupling exactly as in the KSVZ model. Non-perturbative QCD effects then induce the axion potential and dynamically relax the effective strong CP parameter to zero. Again, the mass of the axion satisfies $m_a\propto f_a^{-1}$, for the PQ symmetry-breaking scale $f_a$.

Its cosmological production through vacuum misalignment likewise follows the same general logic as expounded earlier: an initially displaced axion field evolves towards the minimum of its potential, leaving behind a relic abundance that can, for suitable parameters, account for the observed DM density.

To close this discussion, we should say something of hypothesised origins of the axion fields in more fundamental physics. One popular approach is to identify axions as the Kaluza--Klein type modes which arise when one compactifies the extra dimensions in (super)string theory down to the familiar four dimensions of the phenomenal world (see, e.g., \textcite{Svrcek_2006}) (Recall that in Kaluza--Klein compactification, components of higher-dimensional tensor fields appear as scalar fields in lower dimensions upon compactification; see e.g.\ \textcite{Pasini_1988}.)
This compactification, however, doesn't without further inputs fix axion-like particles with exactly the right properties to match the experimental profile of DM (\textcite{Arvanitaki_2010}).

With this background in hand, what to make of axion DM from the point of view of principled inquiry?

\paragraph{Constitutive principles} Much as with SUSY WIMPs, axions don’t infringe on the DM paradigm’s constitutive principles of gravitational physics (except insofar as axions \textit{might} originate in string theory). On the other hand, by augmenting the SM with new particles, both the KSVZ and the DFSZ models modify the SM’s matter sector. From the point of view of the SM, this modification is more localised than that of SUSY WIMPs, though: a surgical intervention, as it were, to insert further particles. The SM’s constitutive principles remain essentially intact. 

The hypothetical physics that axions presuppose is built entirely from the SM's own constitutive toolkit. Everything in the construction of the PQ mechanism---complex scalar fields with Mexican-hat potentials, spontaneous symmetry breaking, Goldstone's theorem, Yukawa couplings, anomalies, EFT integrating-out, etc.---is stock QFT machinery. Renormalisability even functions \textit{actively} as a constitutive selection rule (as the PQ scalar is endowed with the most general renormalisable self-interaction compatible with the assumed symmetries).

That said, the PQ symmetry is a novel constitutive principle, introduced \textit{for the sole purpose of its being anomalously broken}. Absent any independent warrant for it, a principle deliberately engineered to fail at the quantum level is certainly a rather non-standard use of symmetry principles in high-energy physics. 

\medskip

\paragraph{Regulative principles}
\begin{itemize}
\item \emph{Conservativism.} The axion programme introduces new particles and physics, but in a manner that is more \emph{localised} than that of, e.g.\, SUSY WIMPs. As we stressed above, the construction of axionic physics follows standard QFT toolkits. Axions therefore score highly on conservativeness.

That said, the introduction of the PQ symmetry for the sole purpose of its being spontaneously broken is a blemish in this regard. While spontaneously broken approximate global symmetries have ample precedent in particle physics, the particular PQ symmetry and its required anomaly structure have no established counterpart in the pre-axion theory. 

Nevertheless, the charge of \textit{egregious} ad-hocness can be warded off. The axion is deliberately modelled on the pion. Like the latter, it's a pseudo-Goldstone boson associated with a spontaneously broken approximate global symmetry, with its mass generated predominantly by non-perturbative QCD effects. The axion thus imports this familiar structural template into a new symmetry sector. Hence, axions cohere more with our received background than a purely phenomenological \textit{fiat}, but less than the pion itself, since the symmetry-breaking sector responsible for the axion is itself new physics rather than an established feature of QCD.

Which of the two axion models---the KSVZ or the DFSZ model---edges out on conservativeness? On ontological parsimony, the DFSZ model arguably does: it doesn't posit a new heavy quark, but only adds a second Higgs doublet. On non-interference with established structure, the KSVZ model arguably wins: albeit positing more new ontology, it sequesters its new physics and leaves the SM fields' PQ assignments untouched, whereas the DFSZ model posits fewer fundamental fermionic degrees of freedom, but entangles this new physics with existing SM structure. 

\item \emph{Robustness.} 
Axions constitute robust targets for further experimental study due to three features. First, ``(t)he interaction between DM axions […] with particles and forces in the Standard Model leads to a wide variety of ways to search for them” (\cite[p.5]{Chadha-DayAXIONS}). These include, for instance: the conversion of axions into microwave photons inside a resonant cavity immersed in a strong magnetic field (as in haloscope searches); axion-induced effects on electromagnetic fields or particle spins; laboratory photon–axion conversion and re-generation (the `light-shining-through-a-wall’ effect); and astrophysical effects, such as modifications to stellar or supernova evolution resulting from axion production and the consequent energy loss.

Secondly, in comparison to ``the 1980s up to 2010 or so’’ (ibid.), in terms of the experimental technology, ``the landscape has changed’’. Experiments capable of detecting axions are \textit{already running}. The key challenge isn’t detector sensitivity, but rather to search in the right mass regime. The axion mass $m_a$ is broadly unconstrained by theoretical models. Searches must therefore scan mass ranges one narrow slice at a time. The sensitivity required to probe realistic axion models has been achieved already, however: the ADMX experiment, for example, has definitively ruled out axions of the expected strength across several small mass windows \parencite{ADMX:2025}. Moreover, several projects are currently underway specifically to extend that coverage over the next decade (such as MADMAX and similar next-generation detectors aimed at the mass range currently favoured by cosmological simulations, or ALPS II and IAXO for laboratory- and Sun-based experiments). As far as experimental feasibility is concerned, discovery of axions in the near-term future is a live possibility---and ``should reach a conclusion in the next decade or so’’ (\cite[p.11]{Chadha-DayAXIONS}).

Thirdly, what if our instruments find nothing? Would we have only learnt---as in the present case of SUSY WIMPs---that some further stretch of the parameter space is empty \parencite{Graham:2015ouw,Adams:2022pbo}?
At least, in one clear sense axions fare far better than SUSY WIMPs here: the axion's viable parameter space is essentially two-dimensional ($f_a, \theta_i$), versus (say) the Minimally Supersymmetric SM's \emph{124} parameters (cf.\ footnote \ref{fn:MSSM}). As such, exclusions are more meaningful and cumulative in the case of axions, and researchers can hope for better control over ``scanning coverage''.

\item \emph{Generalised Copernicanism.} How typical are axions with the properties required of a viable DM candidate? The answer hinges on \textit{when}—relative to cosmic inflation---the PQ symmetry is broken (see also, e.g., \cite[pp.9]{Chadha-DayAXIONS}). If PQ symmetry breaking occurs \textit{before (or during) inflation}, inflation stretches a single causally connected domain across our observable Universe. It leaves behind the axion with an approximately homogeneous but otherwise arbitrary initial misalignment angle, $\theta_i$. The relic abundance is then sensitive to this initial condition. For sufficiently large $f_a$, the range of values for $\theta_i$ compatible with DM occupies only a fraction of the available parameter space--on some prima facie plausible measures, one in fact tiny in size. To explain such small misalignment angles, one may invoke anthropic selection. This, however, amounts to conceding that our Hubble patch is atypical---precisely the sort of move that Generalised Copernicanism counsels against. 

If PQ symmetry breaking occurs \textit{after inflation}, the initial misalignment angle varies randomly between causally disconnected regions. So the cosmological abundance needn’t depend on a specially selected value of a single $\theta_i$ characterising our entire Hubble patch. The price is that in the post-inflationary scenario one generically has to deal with axion strings, domain walls, and potentially other small-scale structure, all evolving highly nonlinearly. The treatment becomes substantially more complicated and uncertain than the homogeneous misalignment calculation.

The two scenarios therefore involve a trade-off: the pre-inflationary scenario can require an atypical initial condition (especially at large $f_a$), whereas the post-inflationary scenario introduces greater theoretical and computational complexity and uncertainty.

Finally, for more general axion\textit{-like} fields, the space of possible masses, decay constants, and couplings is enormous. Unless a distribution or selection principle can be derived from a more fundamental theory (e.g., from a specified string compactification), selecting an axion-like particle as DM risks appearing \emph{ad hoc}. In this respect, the challenge is again whether the `successful region’ of parameter space is sufficiently typical to satisfy Generalised Copernicanism.




\item \emph{COI inferences.} As axions were originally proposed as part of a mechanism for resolving a distinct, DM-unrelated context---the strong CP problem in QCD---axions exemplify COI-reasoning. Moreover, that the window surviving independent astrophysical and laboratory constraints coincides with the window in which misalignment yields the observed abundance is itself a striking COI argument. The relatively natural emergence of axion-like particles in string theory further strengthened their motivation. (Note, however, that axion-like particles usually cease to be provide solutions to the strong CP problem!)

\end{itemize}

Overall, axions---rightly touted as ``an exceptionally good DM candidate'' (\textcite[p.3]{Marsh2016})---score well with regard to constitutive principles and regulative principles. Their main shortcoming lies in their somewhat compromised standing on Generalised Copernicanism. 

\subsection{Sterile neutrinos}\label{sec:sterileneutrinos}

In its original (`minimal') formulation, the SM predicts neutrinos to be exactly massless. This follows from the minimal SM’s particle content and chiral structure (\textbf{§\ref{ssec:particle-principles}}): it includes only left-handed neutrinos; right-handed neutrinos are absent. The Higgs mechanism (which we may regard here as an integral part of the SM’s electroweak sector) generates the masses of quarks and charged leptons (i.e., electrons, myons, tauons and their neutrinos) through so-called Yukawa interactions between left- and right-handed fermion fields (more on this shortly). Since no right-handed neutrinos exist in the minimal SM, no analogous Yukawa interaction can be written down for neutrinos. The minimal SM's neutrinos are invariably massless.

This prediction clashed with the discovery of neutrino oscillations in the late 1990s. They demonstrated that the neutrino types---the `flavours' electron, muon or tau---produced in weak interactions can transform into one another while propagating. Such oscillations are only possible if the flavour eigenstates participating in weak interactions are quantum superpositions of particles with definite, non-zero masses. Neutrino oscillations, in short, exposed the minimal SM’s incompleteness: the need for its revision in order to bestow masses on neutrinos.

The most conservative emendation enlarges the minimal SM’s particle content by positing \textit{right}-handed neutrino fields. Unlike all other fermions in the SM, these additional fields are gauge singlets: they carry no weak isospin, hypercharge or colour charge; that is, they don’t directly participate in the weak, electromagnetic or strong interactions. Instead, they interact only through gravity and possible mixing with the ordinary neutrinos---hence their appellation, `\textit{sterile} neutrinos'. 

What does positing them purchase us? The heuristic possibilities that open up are again encapsulated in the  Principle of Totalitarianism (cf.\ \textbf{\S\ref{sec:axioms}}): once the particle content and gauge symmetries are specified, one generally writes down every interaction compatible with the fundamental symmetries. By an appeal to the Principle, extra terms can arise for sterile neutrinos: Yukawa coupling terms between left- and right-handed neutrino fields and the Higgs field (entirely analogous to those of the charged leptons and quarks). This Yukawa interaction is the unique renormalisable fermion--Higgs interaction allowed by the SM gauge symmetries between fermions and the Higgs field,  schematically of the form
\begin{equation}
\mathcal{L}_{\text{Yukawa}} \sim y_f \overline{\psi}_L H \psi_R + \text{Hermitian conjugate},
\end{equation}
with $y_f$ the (dimensionless) Yukawa coupling of the fermion $f$, $\psi_L$ and $\psi_R$ are its left- and right-handed components, respectively, and $H$ is the standard Higgs doublet.

When the electroweak symmetry $SU(2)_L \times U(1)_Y$ is spontaneously broken by the Higgs field acquiring a non-zero vacuum expectation value, $\langle H \rangle = v/\sqrt{2}$---that is, when the Higgs field settles into its non-zero ground-state value and acts as a constant background field---these Yukawa interactions generate ordinary (`Dirac') masses for the neutrinos proportional to its Yukawa coupling, 
$m_f = y_f v / \sqrt{2}$, in exactly the same way as for the other fermions. Note that, in order to thus endow neutrinos with mass, the Higgs mechanism needn’t be modified; only the SM’s \textit{field content} in the neutrino sector (i.e., the collection of fundamental fields for neutrinos) is enlarged by including right-handed neutrino fields.

Next, recall that right-handed (sterile) neutrinos are complete singlets under the SM gauge group: that is, they carry no SM gauge charges and don't participate in the SM gauge interactions. Hence, no gauge symmetry forbids them from possessing so-called \textit{Majorana mass terms}. Schematically, they are of the form
\begin{equation}
\mathcal{L}_{\text{Majorana}}
\sim
-\frac{1}{2}M_R\,\overline{N_R^{\,c}}\,N_R
+\text{Hermitian conjugate},
\end{equation}
where $N_R$ denotes a right-handed (sterile) neutrino field, $N_R^{\,c}$ its charge-conjugate field, and $M_R$ the Majorana mass parameter. Unlike a Dirac mass term, which couples distinct left- and right-handed fields, a Majorana mass term couples the sterile neutrino field to its own charge conjugate (i.e., the field obtained by replacing the particle with its corresponding antiparticle). Because the field is coupled directly to its own antiparticle, the resulting particle is identical to its antiparticle. 

Since the Majorana term is already invariant under the SM gauge symmetries, it may be written directly into the Lagrangian (without involving the Higgs mechanism). Accordingly, the Majorana mass parameter $M_R$ is an independent free parameter of the theory; without jeopardising coherence, it may lie at any energy scale.

Majorana mass terms relax a somewhat accidental symmetry, which happens to hold in the SM: conservation of lepton number. While confirmed to high precision, \textit{exact} conservation of lepton number hasn’t been established. By allowing the conversion of neutrinos into anti-neutrinos, Majorana masses violate it. Now,
when both Dirac and Majorana mass terms are present, weak interactions no longer produce \textit{purely} active neutrinos. Instead, the neutrinos created in weak interactions are superpositions of active and sterile fields. 

In the most economical scenario for sterile neutrinos---the so-called canonical Type-I seesaw mechanism (which follows from renormalisability and the postulate of right-handed neutrinos, together with simplicity considerations and the stipulation that the Majorana mass exceed the Dirac mass)---one assumes extremely massive sterile neutrinos (far above the electroweak scale) and allows them to have Majorana masses.

That these Majorana masses are postulated to be very large compared to the electroweak scale is motivated by something like a heuristic Inference to the Best Explanation (cf.\ \textcite{Nyrup_2015}): a large Majorana scale provides a simple and natural explanation for the (phenomenologically given) tiny neutrino masses, without fine-tuning of parameters or complicated field dynamics. More precisely, thanks to the coexistence of large Majorana masses and smaller Dirac masses, we obtain two qualitatively different sets of physical neutrinos. One comprises very heavy and predominantly `quasi-sterile' neutrinos, which interact only extremely weakly with SM particles. The other set comprises very light and predominantly `quasi-active' neutrinos---the neutrinos observed in oscillation experiments. The smallness of the light neutrino masses arises because they are inversely suppressed by the large Majorana mass scale of the heavy sterile states.

Neither the gauge symmetry nor the seesaw mechanism itself fixes the mass scale of the sterile neutrinos. The canonical seesaw picture typically assumes extremely heavy sterile states (with only `natural'---neither extraordinarily large nor small---Yukawa couplings). More general models permit sterile neutrinos with masses spanning many orders of magnitude (including the keV range), with concomitantly adjusted Yukawa couplings. Such particles mix only weakly with the active neutrinos and therefore interact extremely feebly with ordinary matter. Their weak interactions make them cosmologically long-lived, while their masses are sufficiently large for them to behave as non-relativistic matter during structure formation. Under suitable production mechanisms (see \textcite{Merle2017SterileNeutrinoDarkMatter}, whom we closely follow) for generating the observed relic abundance, a keV-scale sterile neutrino exhibits exactly the phenomenological profile of a DM candidate. 

The historically simplest production mechanism is the Dodelson--Widrow mechanism. Its gist is that in the hot plasma of the early Universe, active neutrinos repeatedly scattered from the surrounding particles while simultaneously undergoing ordinary oscillations. Because the sterile neutrino is slightly mixed with the active neutrinos, a small fraction of these oscillations continuously converted active neutrinos into sterile ones. Although each conversion was exceedingly unlikely, the enormous density and duration of the primordial plasma gradually accumulated a cosmologically significant sterile neutrino population.

From a methodological perspective, the Dodelson--Widrow mechanism deserves praise for its economic conservatism: once right-handed neutrinos are admitted, it introduces neither new particles nor new interactions beyond the minimal extension already motivated by neutrino masses. Regrettably, observations ruled out the scenario.\footnote{At least if neutrinos are supposed to provide the dominant component of DM; cf.\ footnote \ref{fn-28} on ``mixed'' approaches.}

More successful proposals compile less economical ingredients. The Shi--Fuller mechanism retains the same particle content but assumes a large primordial lepton asymmetry. Owing to a resonance (which ensues from the changes in the effective neutrino mixing angle), this dramatically enhances the conversion of active into sterile neutrinos at particular energies. The resulting sterile-neutrino population is concentrated at lower momenta than in the Dodelson--Widrow scenario, and is compatible with structure-formation constraints (which spelt doom for the latter). 

Viability of the Shi--Fuller proposal requires that the early Universe possessed a lepton asymmetry several orders of magnitude larger than the observed baryon asymmetry. Some suggestions indeed exist (e.g.,\ the `$\nu$MSM' proposal) for dynamically accounting for such an asymmetry. Yet, neither the lepton asymmetry itself nor---a fortiori---such proposals for accounting for it are independently established.\footnote{That said, fairness mandates that one stress the economic nature of the $\nu$MSM: among Beyond SM theories it’s one of the \textit{least} speculative proposals, adding only three right-handed neutrinos and attempts simultaneously to explain neutrino masses, baryogenesis, and DM. See \textcite{AsakaShaposhnikov2005NuMSM} for details.}

 A third mechanism is `freeze-in production' through the decay of additional heavy particles, such as singlet scalars or inflatons. They populate the sterile-neutrino sector without relying on active/sterile oscillations. Freeze-in scenarios substantially enlarge the viable parameter space and comfortably evade current astrophysical bounds. But they exact the price of postulating new fields or interactions. Admittedly, they are seldom posited \textit{wholly} ad hoc: singlet scalars, inflatons, or other heavy states frequently arise independently in models of inflation, Higgs-portal physics, or ultraviolet completions of the seesaw mechanism---speculative physics in its own right, well beyond established particle physics. Moreover, when employed to produce sterile-neutrino DM, the masses and couplings of those fields are typically selected so as to yield the observed relic abundance, while remaining compatible with cosmological and astrophysical constraints.

\paragraph{Constitutive principles.} Sterile neutrinos leave the SM’s constitutive principles---in particular, the gauge group, Lorentz structure, and all established interaction vertices---untouched. They merely modify the fermionic field content by adding gauge-singlet states: only the particle spectrum in the SM’s neutrino sector is enriched, in a parsimonious and empirically and conceptually well-motivated fashion.
 While the constitutive principles required for sterile neutrinos as DM candidates go beyond minimal extensions of the SM, they plausibly qualify as a `next-to-minimal' extension. Despite to-date unverified status, the physics they implicate remains well-integrated and well-motivated within active particle-physics research, areas of physics that likewise count as reasonable---albeit likewise not minimal---extensions of standard physics. 
\paragraph{Regulative principles.} 
\begin{itemize}
\item \textit{Conservatism}. Sterile neutrinos occupy an ambivalent middle position as DM candidates on Conservatism. They score very high at the level of \textit{particle physics proper}. The introduction of right-handed neutrinos constitutes one of the least invasive extensions of the SM. 

\textit{As DM candidates}, however, sterile neutrinos deviate from conservatism. The requirement that the sterile neutrino mass lie in the keV mass range isn’t dictated by deeper reasons; it’s chosen to fit cosmological and astrophysical constraints. More seriously, the production mechanisms for generating the observed relic abundance rely on additional assumptions beyond the SM’s minimal extensions, which originally motivated sterile neutrinos: either a large primordial lepton asymmetry (in the Shi--Fuller scenario), or the presence of additional heavy degrees of freedom mediating freeze-in production. The former seems a brute fact stipulation for which little independent motivation is forthcoming---even though in itself fairly conservative (but yet unverified) deeper dynamical models exist. The physics of freeze-in production is likewise so-far unverified, and hinges on somewhat speculative ideas. 
\item \emph{Generalised Copernicanism.} Relatedly, the sterile neutrino programme displays a noticeable tension with Copernicanism. The Shi--Fuller production’s primordial lepton asymmetry is many orders of magnitude larger than the baryon asymmetry---a somewhat anti-Copernican hypothesis \textit{at the level of initial conditions}. Freeze-in scenarios presuppose additional heavy particles and corresponding fields whose properties aren’t constrained by terrestrial experiments. The tension is somewhat eased by the coherence with established physics (see above).
\item \emph{COI inferences.} Sterile neutrinos score highly---albeit not perfectly---on COIs.\footnote{E.g., \textcite[p.\ 10]{BoyarskyEtAl2019SterileNeutrinoDarkMatter} expressly endorse COI-reasoning as an overarching regulative principle: ``one tries to minimise the number of new entities introduced but maximise the number of problems which can be addressed simultaneously.”} Right-handed neutrinos were originally motivated by the independent problem of neutrino oscillations. In economical frameworks such as the $\nu$MSM, they even promise a threefold explanatory unification, simultaneously addressing neutrino masses, baryogenesis, and DM. Yet the COI credentials of sterile neutrinos aren’t unqualified. Neither the seesaw mechanism nor the evidence for neutrino masses singles out the keV mass range required for DM, and viable production mechanisms introduce further assumptions that so far aren’t convincingly warranted on independent terms (e.g.\ large primordial lepton asymmetries or additional heavy states).
\item \emph{Robustness.} Embedded in a vibrant and multifaceted experimental programme in neutrino physics (see, e.g.,\ \textcite[chs.\ 7--8]{Merle2017SterileNeutrinoDarkMatter}), work on sterile neutrinos exemplarily aims at robust targets---``a very testable DM candidate” \parencite[p.\ 35]{BoyarskyEtAl2019SterileNeutrinoDarkMatter}. On the cosmological and astrophysical side, its parameter space is constrained by a diverse set of probes (including phase-space limits, X-ray searches for radiative decays, dark-matter abundance constraints, or structure formation). Astrophysical phenomena such as pulsar kicks and supernova dynamics continue to motivate further investigation, illustrating the close connection between sterile neutrino dark matter and cutting-edge problems in contemporary astrophysics.
Complementing these cosmological and astrophysical efforts is an active laboratory programme, with plenty of existing and proposed experiments (including $\beta$-decay spectroscopy, electron-capture measurements, and novel sterile-neutrino capture concepts). 
\end{itemize}

In conclusion, we judge sterile neutrinos to be DM candidates excellently suited to the epistemic situation of the DM problem, with good prospects for testability. While hopes of minimal extensions of the SM to accommodate for sterile neutrinos as DM candidates have been dashed, viable options continue to exist at reasonable costs of conservatism.  Especially noteworthy with respect to principled inquiry is that sterile neutrinos only require an extension of the neutrino ontology. Arguably, this renders them an exceptionally parsimonious proposal within the DM paradigm.\footnote{Research on sterile neutrinos deserves credit for forming part of the first sustained research tradition that sought to identify \textit{standard} neutrinos with DM (e.g.,\ \textcite[ch.\ 6]{Sanders2010}). While those initial efforts failed, resulting in the conclusive refutation of hot DM candidates, this research programme was extraordinarily fruitful and set the standards of all subsequent mainstream DM research, including the WIMP paradigm \parencite{deSwart2025LosingDarkMatter}.}

\subsection{Primordial black holes}\label{sec:PBH}

Primordial black holes (PBHs) are hypothetical relicts from density fluctuations within the first second after the Big Bang (rather than stellar remnants). If these density fluctuations were sufficiently large, gravity would overcome the pressure of the then-dominant radiation and collapse.\footnote{This is the ``most popular, and arguably minimal, PBH formation mechanism” \parencite[p.\ 1]{Green2024PBHBriefOverview}. Other mechanisms exist (e.g.,\ \textcite{CarrKuehnel2022PBHDarkMatterCandidates, CarrGreen2025HistoryPBH}). But they require speculative physics (e.g., cosmic string loops or bubble collisions) and fine-tuning to produce suitable PBHs.} 
By construction, PBHs satisfy the three main features of the DM profile (\textbf{§\ref{sec:DMparadigm}}):

\begin{itemize}
\item \textit{Non-baryonic:} PBHs are created before the QCD phase transition, when free quarks and gluons became confined into hadrons (mainly protons and neutrons). PBHs hence aren’t made up of ordinary matter.
\item \textit{`Cold':} Lest they evaporate via Hawking radiation within the Universe’s present age, the initial mass of PBHs must exceed $10^{-19}$ Solar masses (roughly the mass of a mountain). These cosmologically stable PBHs are too heavy to be relativistic (`hot').
\item \textit{Predominantly gravitional effect:} If we set aside negligible Hawking radiation (in tandem with other conceivable---but uncertain---quantum gravitational effects), PBHs interact with their environment exclusively via gravity. In particular, qua black holes, PBHs are non-luminous.
\end{itemize}
In light of the protracted null-results of experimental searches for DM particles, the main appeal of PBHs is easily grasped: ``(u)nlike most other dark matter candidates, PBHs are not a new elementary particle” \parencite[p.\ 2]{Green2025NonParticleDarkMatter}. 
“However [...] their formation does typically require ‘Beyond the Standard Model’ physics” (ibid.). In fact, \textit{the right amount} of PBHs foists on us hefty ad-hocness: 
\begin{itemize}
\item For the collapse mechanism ``to achieve an interesting (i.e.\ neither negligible nor unphysically large [...]) abundance of PBHs", density fluctuations on smaller scales are needed ``7 \textit{orders of magnitude} larger than its measured value on cosmological scales” \parencite[p.\ 9]{GreenKavanagh2021PBHDarkMatter}. This contrast of mass densities at different scales smacks of ``fine-tuning” (ibid.). 
\item Such a difference in density fluctuations rubs against standard inflation. Today, inflation is, first and foremost, regarded as a promising framework for accounting for structure formation (e.g.,\ \textcite{Smeenk2018InflationOriginsStructure}). How can inflation seed not only galaxies, etc.\ (as per usual), but also---on non-cosmic scales---PBHs? Standard models are nearly `scale-invariant' (i.e., roughly \textit{similar} on cosmic and smaller scales)---in glaring contradiction to what suitable PBH abundances mandate. Scale invariance is ``a generic feature of inflation”: ``the vast majority of inflation models do not generate large PBH-forming perturbations” \parencite[p.\ 8]{Green2025NonParticleDarkMatter}. It’s possible to ``design” \parencite[p.\ 11]{GreenKavanagh2021PBHDarkMatter} inflationary models that reconcile both tasks. But again we must either ``fine-tune” inflation \parencite[p.\ 8]{Green2025NonParticleDarkMatter}, or opt for non-standard models with multiple inflationary fields.    
\item ``Current constraints exclude PBHs from making up all of the DM, apart from in the ‘asteroid mass window’ ($10^{ 17} g \lesssim M \lesssim 10^{22}g$)” (op.cit.,\ p.\ 11). This still permissible range isn’t motivated by independent principles or other considerations. Instead, it’s a brute fact whittling down due to data---last retreats in an ever-shrinking parameter space: that the asteroid-mass region remains open ``largely reflects the difficulty of detecting such light compact objects” \parencite[p.\ 23]{GreenKavanagh2021PBHDarkMatter}.\footnote{Although it's possible give this a more sanguine spin---``[n]ew techniques are required to probe all of the asteroid mass window” \parencite[p.\ 11]{Green2025NonParticleDarkMatter}---one is reminded of what \textcite{Lakatos1989MethodologyResearchProgrammes} scathed as ``degenerative problem-shifts".}     
\end{itemize}

\noindent Appraising PBHs against the standards of the DM paradigm aggravates this unpropitious verdict:

\paragraph{Constitutive principles.} If one grants suitable primordial mass fluctuations as initial conditions, PBHs only require GR. This frugality is marred, however, as soon as one seeks to account for these initial conditions: as we saw above, the requisite physics becomes increasingly exotic and contrived, even by the lights of Beyond SM physics (in the form of typical inflationary scenarios). Production mechanisms for PBHs as a viable DM candidate stray substantially from the SM’s constitutive principles---more so than SUSY WIMPs, axions, or sterile neutrinos.    

\paragraph{Regulative principles.}

\begin{itemize}
\item \emph{COI inferences.} Like the other DM candidates of the paradigm, PBH research betrays a motivation to invoke COI reasoning. Historically, the peak in interest in PBHs as DM candidates was triggered by microlensing results in the early 1990s, and later on again the LIGO-Virgo announcement of gravitational wave discoveries in 2016. The findings \textit{appeared} in conflict with massive stellar sources, suggesting instead primordial ones (\textcite[p.\ 2]{Green2025NonParticleDarkMatter}; see also \textcite{CarrGreen2025HistoryPBH} for details). After subsequent re-examinations and new measurements, a consensus materialised that astrophysical black holes \textit{can} account for the phenomena in question after all (op.cit., p.\ 7).\footnote{\textcite[p.\ 37]{CarrKuehnel2022PBHDarkMatterCandidates}, for instance, write: PBHs ``have been invoked for three main purposes: (1) to explain the dark matter; (2) to generate the observed LIGO/Virgo coalescences; (3) to provide seeds for the SMBHs in galactic nuclei. However, the discussion in Section 4.2 suggests that they could also explain several other observational conundra.'' Because the phenomena, alluded to under the umbrella (2) and (3), are subject to controversy as in need of explanation, and also because the ability of PBHs to actually explain them suffers from great uncertainties and controversy, we’ll refrain here from further discussion.} 

\item \emph{Robustness}. The PBH approach largely flouts the focus on robust targets. The models engineered to achieve the right PBH abundancy make physical assumptions far from theoretical or empirical control. Potential observational effects, at this stage, seem inconclusive: in part because they involve large uncertainties, with ``whether or not the assumptions made in calculating these constraints [...] are reliable (being) a key question'' \parencite[p.\ 11]{Green2025NonParticleDarkMatter}, and in part because ``there may be other potential explanations of these phenomena, and a PBH interpretation may be inconsistent with exclusion limits from other observations” (op.cit., p.\ 9).   

\item \emph{Conservativeness}. Quite generally, and as a corollary of the ad-hocness charges rehearsed above, it’s fair to say that PBH research contravenes the epistemic caution and conservatism, apposite to the DM problem (\textbf{§\ref{ssec:regulative-principles}}): viability comes at the cost of diminished testability, and increasing reliance on speculation and uncertainty (exacerbated by fine-tuning) beyond  independent control or even physical motivation. 
\end{itemize}

In sum, despite initial appearances, we judge PBHs to be close to the edge of credibility within the DM paradigm, and moreover also in isolation a compromisingly ad hoc proposal.

\section{Conclusion}\label{sec:close}

The first of our paper's key findings is quite general.  We articulated a functionalist account of physical principles. It defines principles by the functional roles that hypotheses play in theorising. We distinguished between two broad groups of principles: constitutive principles function as building blocks with a more foundational role, delineating the framework of physical possibilities within which more specific theories and models are sought; regulative principles, on the other hand, order and direct research, their role being akin to methodological maxims and more abstract aims. A principle is justified by reasons that show it to be adequate for that purpose (relative to a particular epistemic situation). Ranging from empirical to pragmatic considerations, such reasons are context-sensitive, depend on (historically variable) background knowledge, and often admit of a certain degree of rational disagreement. 

Principles come into their own in a widespread and powerful (but of course fallible) research strategy for theory or model construction---``principled inquiry”. Rather than proceeding either purely phenomenologically or by unfettered speculation, principled inquirers steer a middle course: principles and the desire to preserve as many and as much of them as possible guide their efforts to advance science. 

Applied to the DM problem, our account revealed that contemporary DM research isn't merely a competition amongst isolated and empirically underdetermined hypotheses. It's more insightfully characterised as the concerted investigation of a theory space, delimited by certain principles.

The lens of principled inquiry illuminates what four of the most popular and mainstream mainstream DM candidates---SUSY WIMPs, sterile neutrinos, axions, and PBHs---have in common, despite their heterogeneity: they can be understood as explorations of the DM paradigm, which largely shares several constitutive and regulative principles. These comprise, in the main, the principles undergirding the general-relativistic framework and the SM. We explicated the reasons that gave the DM paradigm its normative force as a rationally compelling research strategy adapted to the DM problem's epistemic challenges. 

The perspective of principled inquiry also helped us comprehend the despair and ``growing sense of crisis” increasingly aired by DM researchers: the principled approach to the DM problem---exemplified by the foregoing four mainstream candidates---hasn’t delivered the hoped-for success. Instead, it appears that the heuristic resources of principled inquiry into DM are petering out. Pressure is mounting to explore more daunting paths: theoreticians must start bracing themselves for modifying or even jettisoning prima facie well-motivated principles.  

 Nevertheless, even though the crisis is serious, the DM paradigm's resources haven't yet been fully exhausted. The focus on principles induces a natural evaluative matrix---a maxim of minimal mutilation with respect to its principles: the more a DM proposal conforms to the paradigm’s principles (and, conversely, the smaller the requisite deviations from them) the more prioritisation it plausibly deserves. 

This lens yielded a nuanced verdict for prioritising the mainstream DM proposals: while the viability of PBHs and SUSY WIMPs is reaching the limits of \textit{credible} principled inquiry, axions and sterile neutrinos fare significantly better. Both essentially comport with all the principles of the DM paradigm. Sterile neutrinos merely enlarge the SM’s particle spectrum via a new species of neutrinos. They form a well-motivated, and fairly minimal ontological enlargement of the SM. Axions, by contradistinction, involve both a structural and ontological extension: for their introduction, one postulates an additional global symmetry, alongside a scalar field required for its spontaneous breaking. That symmetry postulate, however, has so far received no independent justification, empirical or otherwise. By the standards of the paradigm, we therefore conclude, sterile neutrinos emerge as the most promising of the remaining mainstream candidates.

\section*{Acknowledgements}

We are very grateful to Yemima Ben-Menahem (Hebrew University of Jerusalem), David Kaiser (MIT), Jocelyn Monroe (University of Oxford), and Will Wolf (University of Oxford) for valuable feedback.

%

\appendix

\section{List of abbreviations used}\label{app:abbreviations}

{\small
\begin{description}[
    leftmargin=8em,
    labelwidth=7.5em,
    labelsep=0.5em,
    font=\normalfont\bfseries,
    itemsep=0.15em,
    style=multiline
]

\item[BBN] Big Bang nucleosynthesis
\item[CDM] cold dark matter
\item[CMB] cosmic microwave background
\item[COI] Common Origin Inference
\item[CP] charge conjugation--parity (symmetry)
\item[DFSZ] Dine--Fischler--Srednicki--Zhitnitsky (axion model)
\item[DM] dark matter; also used as the label for the Dark Matter paradigm, i.e.\ the approach that retains GR and posits novel matter
\item[GR] General Relativity
\item[KSVZ] Kim--Shifman--Vainshtein--Zakharov (axion model)
\item[LSP] lightest supersymmetric particle
\item[MOD-GRAV] Modified Gravity, i.e.\ the approach to the DM problem that admits deviations from GR
\item[$\nu$MSM] neutrino Minimal Standard Model
\item[PBH] primordial black hole
\item[PQ] Peccei--Quinn
\item[QCD] quantum chromodynamics
\item[QFT] quantum field theory
\item[SM] Standard Model of Particle Physics
\item[SMBH] supermassive black hole
\item[SR] Special Relativity
\item[SUSY] supersymmetry
\item[UV] ultraviolet (as in `UV-completion')
\item[WIMP] Weakly Interacting Massive Particle
\item[$\Lambda$CDM] Lambda Cold Dark Matter (the standard cosmological model)

\end{description}
}

\printbibliography

\end{document}